\documentclass[showpacs,amsmath,amssymb,aps,10pt,reprint,superscriptaddress]{revtex4-1}
\usepackage{graphicx}
\usepackage{bm}
\usepackage[breaklinks=true,colorlinks=true,linkcolor=blue,urlcolor=blue,citecolor=blue]{hyperref}
\usepackage{graphics}
\usepackage{dcolumn}
\usepackage{amsmath,amssymb}
\usepackage{mathdots}
\usepackage{natbib}
\usepackage{soul,color}
\usepackage{epstopdf}
\usepackage{wasysym,txfonts}

\begin{document}

\title{Mobile fluxons as coherent probes of periodic pinning in superconductors}

\author{Oleksandr V. Dobrovolskiy}
\email[Corresponding author: ]{Dobrovolskiy@Physik.uni-frankfurt.de}
\affiliation{Physikalisches Institut, Goethe University, Frankfurt am Main, 60438, Germany}
\author{Michael Huth}
\affiliation{Physikalisches Institut, Goethe University, Frankfurt am Main, 60438, Germany}
\author{Valerij A. Shklovskij}
\affiliation{Physics Department, V. Karazin National University, Kharkiv, 61022, Ukraine}
\author{Ruslan V. Vovk}
\affiliation{Physics Department, V. Karazin National University, Kharkiv, 61022, Ukraine}

\begin{abstract}
The interaction of (quasi)particles with a periodic potential arises in various domains of science and engineering, such as solid-state physics, chemical physics, and communication theory. An attractive test ground to investigate this interaction is represented by superconductors with artificial pinning sites, where magnetic flux quanta (Abrikosov vortices) interact with the pinning potential $U(r) = U(r + R)$ induced by a nanostructure. At a combination of microwave and dc currents, fluxons act as mobile probes of $U(r)$: The ac component shakes the fluxons in the vicinity of their equilibrium points which are unequivocally determined by the local pinning force counterbalanced by the Lorentz force induced by the dc current, linked to the curvature of $U(r)$ which can then be used for a successful fitting of the voltage responses. A good correlation of the deduced dependences $U(r)$ with the cross sections of the nanostructures points to that pinning is primarily caused by vortex length reduction. Our findings pave a new route to a non-destructive evaluation of periodic pinning in superconductor thin films. The approach should also apply to a broad class of systems whose evolution in time can be described by the coherent motion of (quasi)particles in a periodic potential.
\end{abstract}

\maketitle

\section{Introduction}

The interaction of (quasi)particles with a periodic potential arises in various domains of science and engineering, such as solid-state physics, chemical physics, and communication theory. Exemplary systems include Josephson junctions~\cite{Bar82boo}, superionic conductors~\cite{Ful75prl}, ring laser gyroscopes~\cite{Cho85rmp}, the dynamics of spin \cite{Bar93prl} and charge-density~\cite{Zyb13prb} waves, phase-locking loops~\cite{Ris89boo} in radioengineering, the motion of domain walls \cite{Per08prl}, the magnetization dynamics of interacting spins~\cite{Tit05prb} and the diffusion of colloidal particles in periodic structures~\cite{Evs08pre}. An interesting analogy can be found in superconductivity: Type II superconductors with periodic pinning sites substantiate an attractive test ground to investigate this interaction, thus allowing one to study dynamic properties of the vortex system and scrutinize its force-velocity (current-voltage) response.

The symmetry, shape and intensity of the pinning potential strongly affect the dynamics of Abrikosov vortices in superconductors, as reflected in both, the dc resistance and the microwave (mw) loss \cite{Bra95rpp}. In this regard, tailoring periodic pinning sites has been revealed to be most efficient for controlling the resistive response of superconductors \cite{Plo09tas,Sol14prb,Pry06apl}, dynamic mode-locking \cite{Jel15nsr}, stimulation of superconductivity at microwave frequencies \cite{Lar15nsr}, enhancing speed limits for the dynamics of Abrikosov vortices \cite{Sil12njp,Gri15prb,Shk17prb,Dob17sst} as well as enabling functionality of such fluxonic devices as rectifiers \cite{Vil03sci}, transistors \cite{Vla16nsr}, and high-frequency filters \cite{Dob15apl}. The availability of high-resolution nanofabrication tools has advanced the use of more and more sophisticated, asymmetric pinning geometries (vortex ratchets) \cite{Plo09tas,Vil03sci}. For instance, in a vortex ratchet device (diode), the pinning site array may consist of elements of different sizes and shapes, such as triangles \cite{Vil03sci}, grading circles \cite{Gil07prl}, and arrow-shaped wedged cages \cite{Tog05prl}. The coordinate dependence of the resulting pinning potential $U(x,y)$ in these is rather complex, which is why for a quantitative interpretation of data an assumption is usually made for $U(x,y)$ a priori. In the case of a simpler washboard pinning potential (WPP) \cite{Dob17pcs}, $U(x,y)\equiv U(x) = U(x+a)$, where $a$ is the period, an asymmetry can be induced by pre-defining the steepness of the left and right slopes of linear-extended pinning ``sites'' differently. The respective 3D washboard ratchet nanostructures can be realized, e.g., by milling of uniaxial nanogrooves by a focused ion beam (FIB) into the film \cite{Dob15met}, or via ferromagnetic decoration of the films by focused electron beam induced deposition (FEBID) \cite{Dob10sst}. Since the geometrical appearance of an asymmetric washboard nanostructure does not reveal the asymmetry and the shape of the resulting WPP, non-destructive approaches for the determination of $U(x,y)$ are needed.

The ideas for the determination of $U(x,y)$ from experiment have been attracting attention of researchers for half a century \cite{Cam72aip,Low72jpf,Aus09nph,Emb15nsr,Kre16nal}. Early approaches for the determination of the distribution, density and strength of pinning sites included proposals to quantify the pinning force from measurements implying a small ripple magnetic field superimposed on a larger dc magnetic field~\cite{Cam72aip} and from measurements with a small ac current \cite{Low72jpf}. Recently, elastic properties of individual vortices were investigated by magnetic force
microscopy \cite{Aus09nph}. Further, scanning SQUID microscopy has been used to probe the dynamics and pinning of single vortices under combined dc and small ac drives, and the dependence of the elementary pinning force of multiple defects on the vortex displacement has been measured \cite{Emb15nsr}. It has also been shown \cite{Kre16nal} that vortices respond to local mechanical stress applied in the vicinity of a vortex thus allowing one to manipulate individual vortices without magnetic field or current. In all these experiments the samples contained randomly distributed pinning sites and either the vortex response in a non-coherent regime \cite{Cam72aip,Low72jpf} or that of an individual vortex \cite{Aus09nph,Emb15nsr,Kre16nal} was probed.

An intriguing situation ensues in a washboard pinning landscape at a particular value of the magnetic field $H_m$ when each row of vortices is pinned at the bottom of the linearly-extended pinning sites (nanogrooves) and there are neither vacant nanogrooves nor vortices pinned between them. This field $H_m$ is the fundamental matching field at which the vortex lattice is commensurate with the pinning landscape. Computer simulations revealed \cite{Luq07prb} that in this case the vortex lattice has a crystalline structure, the effective intervortex interaction is cancelled, and each vortex experiences the same pinning force. An important consequence emerges from this: In the coherent regime, the moving ensemble of vortices in a WPP can be regarded as a moving vortex crystal, thus allowing one to interpret data employing single-vortex mechanistic models \cite{Git66prl,Cof91prl,Pom08prb,Shk08prb,Shk11prb}.

Here, in contradistinction to previous works \cite{Cam72aip,Low72jpf,Aus09nph,Emb15nsr,Kre16nal}, we use the coherent vortex dynamics at the fundamental matching field to examine a theoretical mechanistic approach \cite{Shk12inb} for the determination of the coordinate dependence of a periodic pinning potential in superconductors under combined dc and mw current stimuli. Specifically, we evaluate the dc current-induced reduction of the depinning frequency in nanopatterned Nb films with different groove slope asymmetries and determine the dependences $U(x)$ from ensemble-integrated microwave power absorption data. Further, using the pinning asymmetry parameters thus deduced, we augment the validity of the presented approach by a good fitting of mode-locking steps in the electrical voltage response in the presence of an ac drive to expressions derived within the framework of a stochastic model \cite{Shk14pcm} of anisotropic pinning.

\section{Results}
\begin{figure}[t!]
    \centering
    \includegraphics[width=1\linewidth]{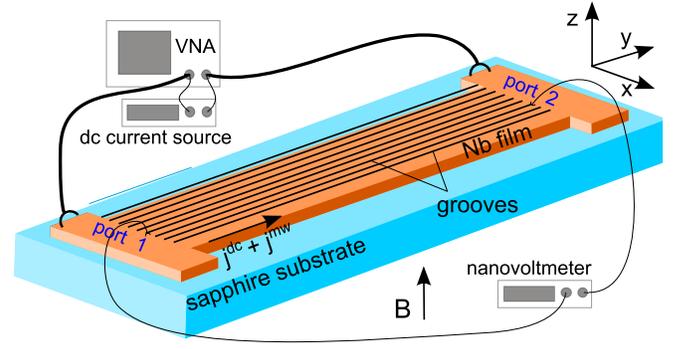}
    \caption{Experimental geometry: A combination of dc and mw current densities $j+j^{mw}$ is applied to a superconducting Nb microstrip. The microstrip contains an array of $500$\,nm-spaced nanogrooves milled by focused ion beam. The Lorentz force causes the vortices to move along the $x$-axis. The measurable quantities are the dc voltage and the absolute value of the forward transmission coefficient $S_{21}$ of the mw power at port 2 with respect to that at port 1.}
    \label{fGeom}
\end{figure}

We studied the vortex-state resistive response of Nb microstrips by combined microwave transmission spectroscopy and electrical voltage measurements, Fig. \ref{fGeom}. The $500\,\mu$m long and $150$\,$\mu$m wide microstrips were patterned with arrays of $500$\,nm-spaced nanogrooves milled by FIB. The nanogrooves were parallel to the direction of the transport current density $\mathbf{j}$ which, in a perpendicular magnetic field $\mathbf{B} = \mu_0 \mathbf{H}$, exerted a transverse Lorentz force $\mathbf{F}_L = \mathbf{j}\times \mathbf{B}$ on vortices leading to their dissipative motion in the direction \emph{perpendicular to the grooves}. The nanogroove arrays induced a periodic pinning potential of the washboard type $U(x,y)\equiv U(x)= U(x+a)$. Sample S was patterned with nanogrooves having a symmetric cross-section, sample A1 with nanogrooves having a weak asymmetry of the groove slopes, and the control sample A2 with nanogrooves having a strong asymmetry of the slopes. Atomic force microscope images of the nanogrooves are presented further in the Discussion section. The asymmetric grooves are oriented in such a way that a positive dc bias makes the vortices to probe the gentle slope of the WPP: The vortices are shaken at the bottom of the pinning potential, which is gradually shifted in the negative $x$ direction as the dc bias magnitude is increased. Samples S, A1, and A2 are $40$\,nm, $56$\,nm and 70\,nm thick and have a superconducting transition temperature $T_c$ of $8.66$\,K, $8.85$\,K, and $8.94$\,K, respectively. The upper critical field for all samples at zero temperature is estimated as $H_{c2}(0) = H_{c2}(T)/ [1-(T/T_c)^2]\approx 1$\,T corresponding to a coherence length of $\xi(0) = [\Phi_0/2\pi H_{c2}(0)]^{1/2} \approx 17$\,nm. The diameter of the vortex core $\simeq2\xi$ at $T = 0.3 T_c$ is by a factor of 6 smaller than the full width at half depth of the grooves in samples A1 and A2, and by a factor of 3 for sample S.

\subsection{``Ratchet window'' in microwave power absorption}

\begin{figure}[b!]
    \centering
    \includegraphics[width=1\linewidth]{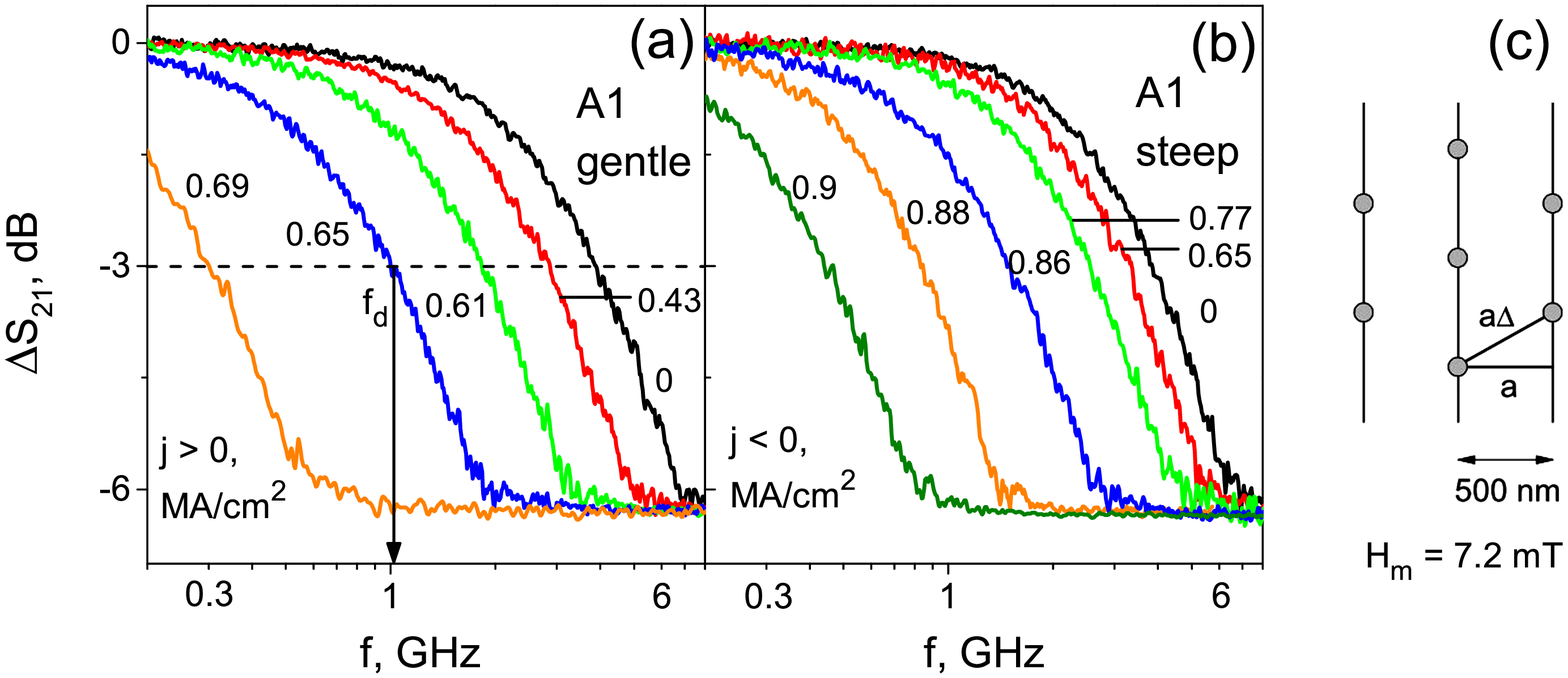}
    \caption{Frequency dependence of the relative change of the absolute value of the forward transmission coefficient $\Delta S_{21}(f)$ of sample A1 at positive (a) and negative (b) dc densities, as indicated. The arrow in (a) indicates the definition of the depinning frequency $f_d$ by the $-3$\,dB criterion. (c) Vortex lattice configuration with lattice parameter $a_\bigtriangleup = (2\Phi_0/H\sqrt{3})^{1/2}$ and the matching condition $a_\bigtriangleup = 2a/\sqrt{3}$ in a washboard nanolandscape with period $a = 500$\,nm at the fundamental matching field $H_m = 7.2$\,mT.}
    \label{fSvF}
\end{figure}

We investigated the frequency dependence of the mw power absorbed by vortices in Nb microstrips with symmetric and asymmetric nanogrooves at the fundamental matching field $H_m = 7.2$\,mT and $T/T_c = 0.3$ for a series of dc bias values of both polarities. The data for the absorbed mw power were acquired in terms of the relative change of the absolute value of the forward transmission coefficient $\Delta S_{21}(f) \equiv S_{21} - S_{21\mathrm{ref}}$, where $S_{21\mathrm{ref}}$ is the reference mw loss in the transmission line (all cables, connectors etc.) and $\Delta S_{21}$ is a measure for the mw loss due to vortex motion in the sample under study, Fig. \ref{fSvF}. Here all quantities are expressed in dB since it is the \emph{ratio} of the transmission coefficients with and without the microstrip device which is a quantity of our interest.

\begin{figure*}[t!]
    \centering
    \includegraphics[width=1\linewidth]{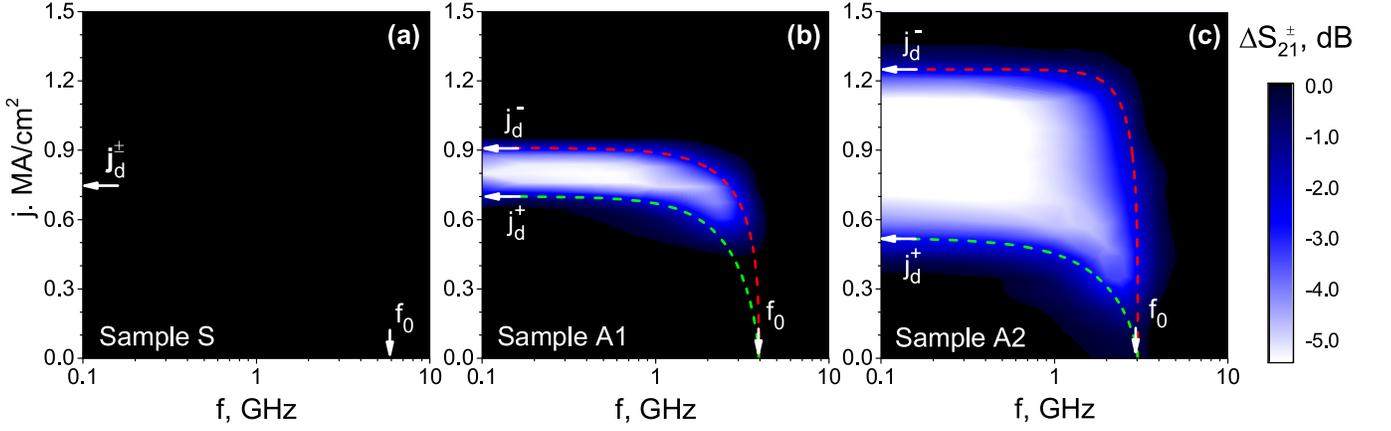}
    \caption{Difference signals $\Delta S_{21}^{\pm} = \Delta S_{21}(j^+) - \Delta S_{21}(j^-)$ for sample S (a), A1 (b), and A2 (c) at $T = 0.3 T_c$ and the fundamental matching field $H_m = 7.2$\,mT. The arrows indicate the zero-bias depinning frequencies $f_d(j=0)$ and the depinning current densities $j_d^{+}$ and $j_d^{-}$ for the motion of vortices against the gentle and steep slope of the grooves, respectively. The dashed lines are fits of the general form $f_d/f_d(j=0) = [1 - (j/j_d)^{k/l}]^{m/n}$, with exponents $k,l,m,n$ indicated in Fig. \ref{fPot}.}
    \label{fScontour}
\end{figure*}

In what follows we focus on the data acquired at a mw excitation power of $P = -20$\,dBm, leaving an analysis of nonlinear effects observed at higher power levels beyond the scope of this work. In the order of magnitude, the amplitude of the vortex displacement at $P = -20$\,dBm at a frequency of $1\,$GHz can be estimated as 10\,nm on the basis of complementary fluctuation spectroscopy measurements on a series of narrow Nb microstrips fabricated from as-grown flat and grooved Nb films. The data at $P = -30$\,dBm lead to essentially the same results but the data are more noisy. Contrarily, further effects emerge at high excitation power levels of $P = -6$\,dBm and $-3$\,dBm in the nonlinear regime which should be analyzed with an in-depth analysis of overheating effects and the associated with them quasiparticle escape from the vortex cores \cite{Lar15nsr}. In this way, in the absence of the dc current, the ac-driven vortex dynamics is substantially in the \emph{linear regime} (vortex displacements are small) and it transits into the dc-induced nonlinear regime as the dc bias is increasing so that the vortices are \emph{shaken while overcoming the WPP barriers}.

Figure \ref{fSvF}(a,b) presents the raw data $\Delta S_{21}(f)$ for sample A1, while the respective data for samples S and A2 are reported in Supplementary. The arrangement of vortices at $7.2$\,mT in the pinning nanolandscape is shown in Fig. \ref{fSvF}(c) for the assumed triangular vortex lattice with lattice parameter $a_\bigtriangleup = (2\Phi_0/B\sqrt{3})^{1/2}$ and the matching condition $a_\bigtriangleup = 2a/\sqrt{3}$.  In Fig. \ref{fSvF} $\Delta S_{21}(f)$ has a smooth crossover from the weakly-dissipative regime at low frequencies ($\Delta S_{21}(f)= 0$ corresponds to the minimal mw loss) to the strongly-dissipative regime at high frequencies ($\Delta S_{21}(f)\approx -6.3$\,dB is the maximum mw excess loss due to vortices).

The crossover takes place at the depinning frequency $f_d$ determined at the point where the phase difference between the viscous and pinning forces amounts to $\pi/2$\cite{Shk12inb} and corresponds to about $-3$\,dB excess loss level in our measurements, which is indicated as a determination criterion for $f_d$ in Fig. \ref{fSvF}(a). At low frequencies the vortices visit many potential wells (delocalized regime), whereas at high frequencies the vortices are shaken at the groove bottoms (localized regime). We note that the pinning force $F_p = -\nabla U(x)$ attains a maximum at the groove slopes, whereas $F_p \approx 0$ in the vicinity of the groove bottoms where $U(x)$ is nearly constant. Accordingly, overcoming of the pinning barriers by vortices at low frequencies causes the pinning force to dominate the viscous force and the response is weakly dissipative, whereas in the high-frequency regime the viscous force dominates the pinning force, thus leading to a strongly dissipative response \cite{Git66prl}. The zero-bias depinning frequency $f_0 \equiv f_d(j=0)$ of samples S, A1 and A2 at $T/T_c=0.3$ and $H_m=7.2$\,mT  amounts to $5.72$, $3.95$, and $3.12$\,GHz, respectively.

For both dc bias polarities $f_d$ shifts towards low frequencies with increasing dc bias value, Fig. \ref{fSvF}(a,b). Under dc polarity reversal, the magnitudes of the $f_d(j)$ shifts noticeably differ for sample A1, even more so for sample A2, but nearly coincide for sample S, see Supplementary. The reduction of the depinning frequency upon increasing the dc bias can be understood as a consequence of the effective lowering of one of the pinning potential wells due to its tilt caused by the dc bias \cite{Shk10ltp}. The mechanistic consideration of a vortex as a particle leads to the conclusion that during an ac semiperiod, while the dc-tilted pinning potential well is broadening \cite{Shk10ltp}, with increasing frequency $f$ the vortex has no longer time $\propto 1/f$ to reach the groove slopes where the pinning forces dominate. Accordingly, the response becomes stronger dissipative already at lower frequencies, as compared with the zero-bias curve.
\begin{figure*}[t!]
    \centering
    \includegraphics[width=1\linewidth]{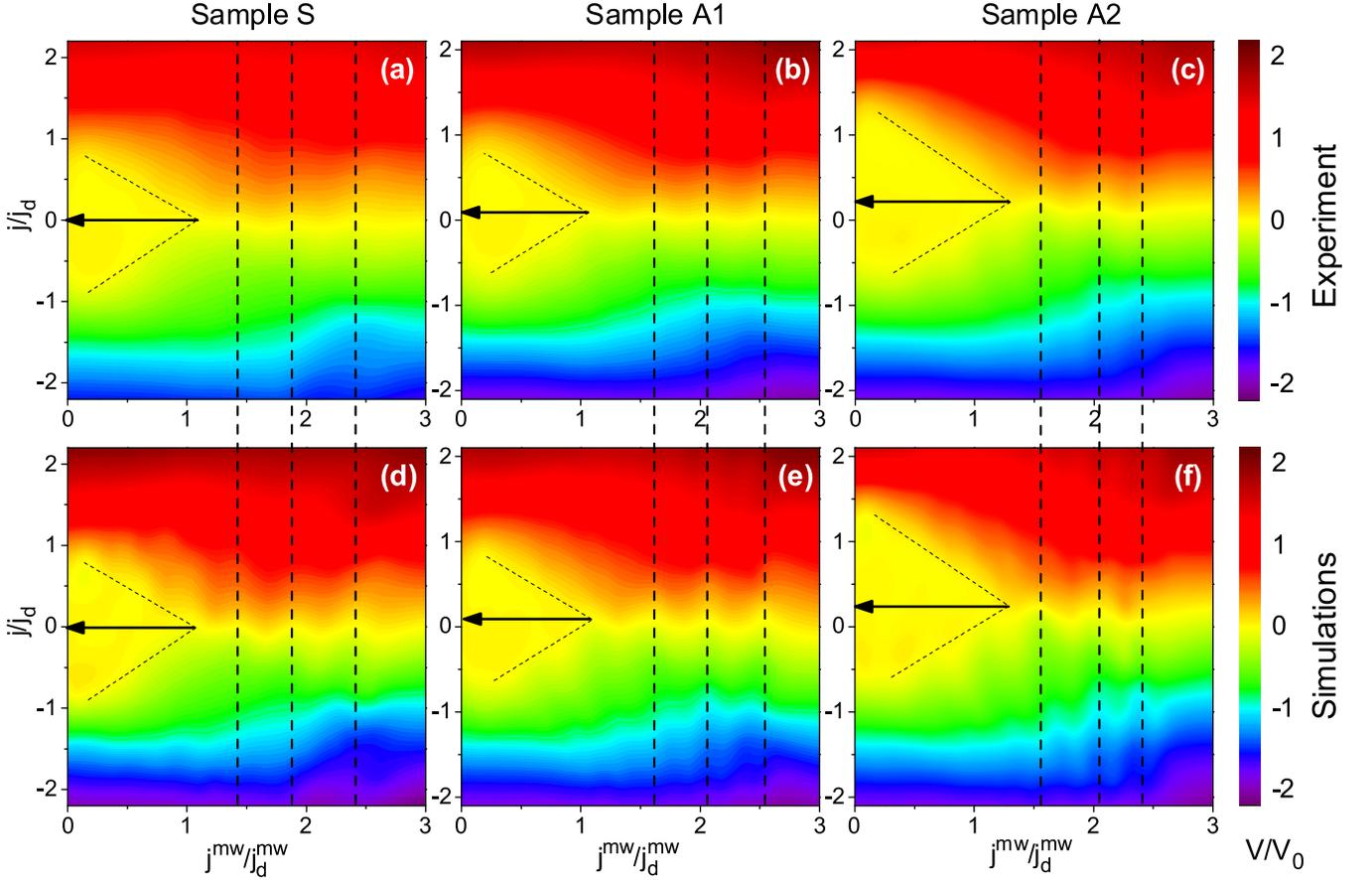}
    \caption{Electrical dc voltage as a function of the normalized dc current density and the microwave amplitude for all samples at $T = 0.98 T_c$ and $H_m = 7.2$\,mT. (a-c) Experimental data. (d-f) Theoretical modeling results, as detailed in Discussion. The color scale is the same in all panels with the electrical voltage normalized to the first Shapiro step voltage $V_0$. Vertical dashed lines mark mode-locking fringes. Arrows depict the dc loading capability of the vortex ratchets, as deduced from crossing the straight lines $j_d(j^{mw})$ at the point where the internal asymmetry of the pinning potential is effectively compensated by the extrinsic asymmetry of the WPP caused by the dc bias.}
    \label{fSteps}
\end{figure*}

The difference in the magnitudes of the current-induced shifts of $f_d$ of samples A1 and A2 is best seen in Fig. \ref{fScontour} displaying contour plots of the difference signal $\Delta S_{21}^{\pm} = \Delta S_{21}(j^+) - \Delta S_{21}(j^-)$ which serves as a measure of the ratchet properties of the samples. The bright areas in the contour plots in Fig. \ref{fScontour}(b) and (c) represent the ``ratchet windows'' in the microwave power absorption, i.e. the frequency - current range where the system exhibits rectifying properties, i.e. works as a diode. As expected, sample S with symmetric grooves does not exhibit ratchet properties, as the absolute value of the depinning current density does not change under dc bias polarity reversal in this sample. The ``ratchet window'' of samples A1 and A2 is bound by the gentle-slope $j^+_d$ and the steep-slope $j^-_d$ depinning current densities (extrapolated to $f\rightarrow 0$ corresponding to the dc regime), as well as by the value of the zero-bias depinning frequency $f_d(j=0)$, as indicated in Fig.\,\ref{fScontour}. The depinning current density $j_d$ amounts to $0.75$\,MA/cm$^2$ for both CVC branches of sample S, $0.91$\,MA/cm$^2$ and $0.7$\,MA/cm$^2$ for the steep-slope and the gentle-slope direction of the vortex motion in sample A1,  and $1.25$\,MA/cm$^2$ and $0.52$\,MA/cm$^2$ for the steep-slope and the gentle-slope direction of the vortex motion in sample A2, respectively. The depinning current densities $j_d^{\pm}$ are determined by the $10\,\mu$V/cm electric field strength criterion from the current-voltage curves (CVCs) discussed next.

\subsection{Mode-locking steps in the current-voltage curves}

\begin{figure}[t!]
    \centering
    \includegraphics[width=0.8\linewidth]{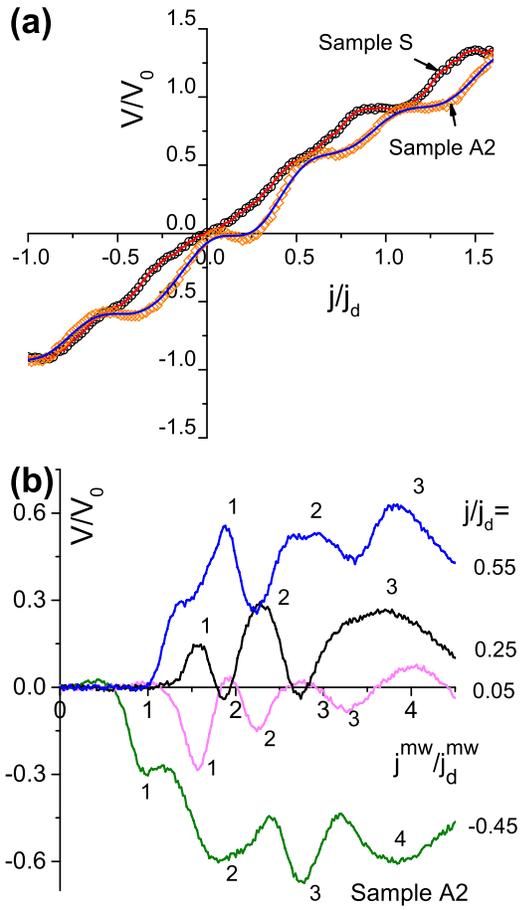}
    \caption{(a) Shapiro steps in the CVC of microstrips S and A2 at $j^{mw}/j^{mw}_d = 1.83$, $T = 0.98T_c$, and the fundamental matching field $H_m = 7.2$\,mT. Symbols: Experiment. Solid lines: Fits to Eq. (\ref{eCVC}). (b) Mode-locking ``fringes'' in the dc voltage with increasing mw amplitude at $T = 0.98T_c$, $H_m = 7.2$\,mT, and a series of dc densities as indicated.}
    \label{fLocking}
\end{figure}
Figure \ref{fSteps}(a-c) displays the dc electrical voltage as a function of the normalized dc density and the ac amplitude at $T = 0.98T_c$ and the fundamental matching field $H_m = 7.2$\,mT. In all panels $f = 0.3 f_d$, with $f_d$ amounting to $444$, $306$, and $242$\,MHz, for samples S, A1, and A2, respectively. In the absence of microwave excitation, the CVC of Sample S in Fig. \ref{fSteps}(a) is symmetric, while the CVCs of Sample A1 and A2 demonstrate two different absolute values of the depinning current density $j_d^+ > |j_d^-|$ for the positive and the negative branch. The addition of the microwave stimulus leads to the appearance of Shapiro steps in the CVC. The steps occur at voltages \cite{Fio71prl} $V = n V_0 \equiv nN\Phi_0 f$, where $n$ is an integer, $N$ is the number of vortex rows between the voltage leads, $f$ is the microwave frequency, and $\Phi_0 = 2.07\times10^{-15}$\,Vs is the magnetic flux quantum. The steps in the CVCs, which are best seen in the conventional current-voltage representation in Fig. \ref{fLocking}(a), arise when the hopping distance of Abrikosov vortices during one ac halfwave coincides with one or a multiple of the nanostructure period. These interference steps are a fingerprint of the coherent vortex dynamics. They remain visible in the field range from $7$ to $7.5$\,mT and disappear as $H$ is tuned further away from the fundamental matching field $H_m = 7.2$\,mT.

We note that temperatures close to $T_c$ are fortunate for the observation of Shapiro steps which are best seen in the flux flow regime at $j> j_d$. This is because of the depinning current decreasing with increasing temperature and the flux flow setting on at smaller dc densities $j_d< j < j^\ast$ thus allowing one to operate in an extended current range where overheating effects are negligible. Here $j^\ast$ is the current density corresponding to an abrupt transition of the sample into the normal state due to the Larkin-Ovchinnikov instability \cite{Lar75etp,Mus80etp,Kle85ltp,Vol92fnt,Per05prb,Sil12njp,Leo16prb,Dob17sst,Shk17prb}.

In Fig. \ref{fSteps}(a-c) one sees that $j_d$ nearly linearly decreases with increasing microwave amplitude and at some point the lines $j_d(j^{mw})$ for different polarities cross, thus allowing one to deduce a dc bias value for which an effective ``symmetrization'' of the WPP is predicted~\cite{Shk14pcm}: At this dc bias value the internal asymmetry of the pinning potential is effectively compensated by the extrinsic asymmetry of the WPP caused by the dc bias. Here we refer to Fig. 2 in Supplementary for a sketch. Quantitatively, these dc values characterize the loading capability of the ratchet \cite{Knu12pre} and amount to $j/j_d\approx0.1$ for sample A1 and $j/j_d\approx0.25$ for sample A2. Obviously, the lines $j^{\pm}_d(j^{mw})$ cross at $j/j_d\approx0$ for sample S as this sample has symmetric grooves. As the ac amplitude increases, the dc voltage for all samples exhibits mode-locking peculiarities (``fringes''). Up to four lowest-order fringes for sample A2 are exemplified in Fig. \ref{fLocking}(b) for a series of sub-depinning dc biases of both polarities. Both, Shapiro steps and mode-locking fringes were theoretically predicted earlier for symmetric \cite{Shk11prb} and asymmetric \cite{Shk14pcm} WPPs. To examine whether the voltage responses can be described theoretically, we determine the coordinate dependence $U(x)$ of the pinning potential from the microwave power absorption and use the deduced expressions for $U(x)$ for modeling of the voltage responses in Fig. \ref{fSteps}(d-f), as detailed next.

\section{Discussion}
\subsection{Determination of the coordinate dependence of the pinning potential}
\begin{figure*}[t!]
    \centering
    \includegraphics[width=1\linewidth]{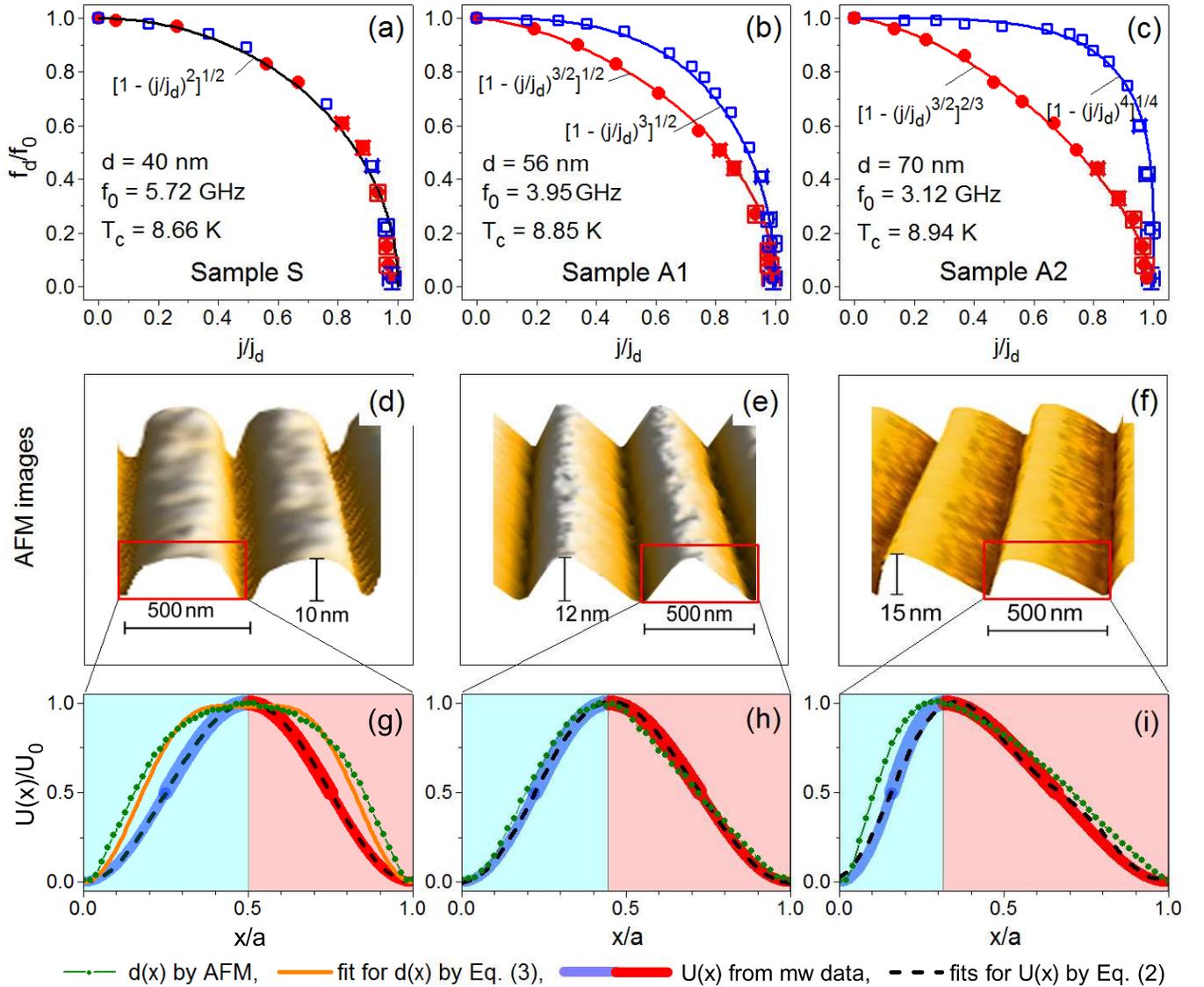}
    \caption{\small\linespread{1.02}\selectfont{}(a)-(c) Reduction of the depinning frequency upon increasing the dc density as deduced from the mw power absorption data.  The experimental data for the positive ($\medbullet$) and the negative ($\square$) dc polarity are approximated by fits (solid lines of respective color) of the general form $f_d/f_0 = [1 - (j/j_d)^{k/l}]^{m/n}$, with the exponents $k,l,m,n$ as indicated. (d)-(f) Atomic force microscope images of the WPP landscapes milled by FIB on the surface of samples S, A1, and A2. When subjected to a combination of microwave and dc currents, fluxons act as movable coherent ``sensors'' probing the curvature of $U(x)$: The ac component shakes the fluxons in the vicinity of their equilibrium points which are unequivocally determined by the local pinning force counterbalanced by the driving (Lorentz) force induced by the dc bias current. A positive dc bias makes the vortices to ``probe'' the gentle groove slope, whereas the steep groove slope is probed by vortices when they are subject to a negative dc bias. (g)-(i) The coordinate dependences of the pinning potentials $U(x)$ deduced numerically using the approach outlined in Supplementary (solid lines) are plotted together with fits (dashed lines) to Eq. \eqref{eDeduced} and the cross-sectional AFM line scans across the grooves (symbols). The light red areas in panels (g)-(i) correspond to the left-hand WPP groove slopes in panels (a)-(c) probed at positive dc bias values (red solid symbols in panels (a)-(c)). The light blue areas in panels (g)-(i) correspond to the right-hand WPP groove slopes in panels (a)-(c) probed at negative dc bias values (blue open symbols in panels (a)-(c)).}
    \label{fPot}
\end{figure*}

Figure \ref{fPot}(a)-(c) displays the dependences $f_d/f_0$ versus $j/j_d$ deduced from the raw experimental data $\Delta S_{21}(f)$, as shown in Fig. \ref{fSvF} for a series of dc bias values of both polarities for all samples. The data acquired for positive dc biases are shown by solid red symbols, while those for negative dc biases by open blue ones. The relative uncertainty in the determination of the depinning frequency does not exceed $7\%$ at $j \rightarrow j_d$ and it is smaller than $3\%$ at dc current densities $j < 0.8j_d$. This is indicated by the error bars in Fig. \ref{fPot}(a)-(c), which become smaller than the symbol size at $j < 0.8j_d$. The larger error in the determination of $f_d$ at $j \rightarrow j_d$ is associated with a smeared functional shape of the curves $\Delta S_{21}(f)$ in Fig. \ref{fSvF} upon vanish of the WPP barriers due to their tilt by the dc current.

To interpret the dc-induced reduction of the depinning frequency $f_d/f_0$ in Fig. \ref{fPot}(a-c) as a function of the dc current density $j/j_d$ at $0.3T_c$ for the coherent vortex at the fundamental matching field $H_m = 7.2$\,mT, we employ a single-vortex theoretical approach outlined in Supplementary. This approach is based on the Gittleman-Rosenblum model \cite{Git66prl} generalized for the presence of a dc current bias \cite{Shk08mmt,Shk13ltp,Shk12inb}. The procedure of determination of the coordinate dependence of the pinning potential from the microwave power absorbed by vortices requires to approximate the reduction of the depinning frequency for both dc polarities by some expression. The experimental data for all our samples fit to expressions of the universal ``mean-field'' type
\begin{equation}
    \label{eFits}
    f_d/f_0  = [1 - (j/j_d)^{k/l}]^{m/n}
\end{equation}
with the exponents $k/l$ and $m/n$ labeled at the respective fit curves in Fig.~\ref{fPot}(a)-(c). For consistency, we also plot the respective fit curves in the contour plots in Fig. \ref{fScontour} and note that they nicely encage the ``ratchet windows'' quantitatively.

The employment of the procedure outlined in Supplementary to the fit curves in Fig.~\ref{fPot}(a)-(c) results in the $U(x)$ curves shown in Fig. \ref{fPot}(g)-(i) by solid lines. The curves $U(x)$ in Fig. \ref{fPot}(g)-(i) are plotted for one period of the pinning potential scaled to its depth $U_0$ and period $a$. The line thickness reflects the uncertainty in the determination of $U(x)$, which does not exceed $10\%$ and $5\%$ on the dome and at the bottom of the potential, respectively. Analytically, the dependences $U(x)$ deduced from the mw data can be approximated by expressions of the general form
\begin{equation}
    \label{eDeduced}
     U(x)/U_0 = [(1 - \cos2\pi x/a) + e (1-\sin4\pi x/a)/2]/2
\end{equation}
with $e$ being the asymmetry parameter and amounting to $0$, $0.13$ and $0.56$ for samples S, A1, and A2, respectively.

Figure \ref{fPot}(g-i) displays the cross-sections of the nanogrooves measured by atomic force microscopy (AFM) in comparison with the dependences $U(x)$ deduced from the mw data and the respective approximations by Eq. \eqref{eDeduced}. Obviously, the deduced coordinate dependence of the pinning potential $U(x)$ in Sample A1 corresponds quite well to the coordinate dependence of the film thickness $d(x)$ modulated by the nanogrooves, as measured by AFM. The AFM profile for Sample A2 can also be fitted satisfactorily to the same expression as $U(x)$. At the same time, $U(x)$ for sample S reproduces the AFM profile qualitatively, but does not follow it quantitatively. The modulated thickness $d(x)$ in Sample S can be fitted to
\begin{equation}
    \label{eAFMsym}
    d(x) = [(1 - \cos2\pi x) + 0.56(1-\cos4\pi x)/2]/2.
\end{equation}

The stronger discrepancy between the dependences $U(x)$ and $d(x)$ in sample S may be caused by a factor of about two smaller groove width in Sample S as compared to Samples A1 and A2. In all, we draw the conclusion that the employed procedure \cite{Shk12inb} is not only sensitive to the asymmetry of the pinning potential, but it also reveals a correlation between $U(x)$ and $d(x)$ thus indicating that pinning in the two thicker samples is primarily caused by vortex length reduction. This can be understood as the vortex energy $\varepsilon(x)$ is proportional to its length $L(x)$ determined by the film thickness $\varepsilon(x) \propto L(x) = d(x)$.

\subsection{Modeling of electrical resistance responses}
\label{sModel}

To augment the validity of the employed approach for the determinations of the coordinate dependences $U(x)$ of the pinning potential from the microwave power absorption at $T = 0.3T_c$ we employ a stochastic model of uniaxial periodic pinning \cite{Shk08prb,Shk11prb,Shk14pcm} to fit the peculiarities observed in electrical voltage responses at noticeably higher temperatures $T = 0.98T_c$. The theoretical treatment relies upon the Langevin equation for a vortex moving with velocity $\mathbf{v}$ in a magnetic field $\mathbf{B}=\mathbf{z}B$, where $B\equiv|\mathbf{B}|$, and $\mathbf{z}$~is the unit vector in the $z$ direction
\begin{equation}
        \label{eLE}
        \eta\mathbf{v} = \mathbf{F}_{L}+\mathbf{F}_{p}+\mathbf{F}_{th},
\end{equation}
where $\mathbf{F}_{L}=(\Phi_{0}/c)\mathbf{j}\times\mathbf{z}$ is the Lorentz force, $\Phi_{0}$ is the magnetic flux quantum, and $c$ is the speed of light. In Eq. \eqref{eLE} $\mathbf{j}\equiv\mathbf{j}(t)= \mathbf{j}+ \mathbf{j}^{mw} \cos\omega t$, where $\mathbf{j}$ and $\mathbf{j}^{mw}$ are the dc and ac current density amplitudes and $\omega$ is the angular frequency. $\mathbf{F}_{p}=-\nabla U_p(x)$ is the anisotropic pinning force, where $U_p(x)$ is a ratchet WPP. $\mathbf{F}_{th}$ is the thermal fluctuation force represented by Gaussian white noise and $\eta$ is the vortex viscosity.

The ratchet WPP is modeled by
\begin{equation}
        \label{eWPP}
        U_p(x) = (U_p/2) [1-\cos kx + e(1 -\sin 2kx)/2],
\end{equation}
where $k=2\pi/a$ with $a$ being the period and $U_p$ the depth of the WPP. In Eq. \eqref{eWPP} $e$ is the asymmetry parameter allowing for tuning the asymmetry strength. It is this parameter which must be determined from experiment.

The asymmetry parameter deduced from the experimental data amounts to $e = 0$ for Sample S, $e = 0.13$ for Sample A1, and $e = 0.56$ for Sample A2. These values of $e$ are used in the simulations of the electric field response given by \cite{Shk14pcm}
\begin{equation}
    \label{eCVC}
    E= \nu(j, j^{mw}, f, t)j.
\end{equation}
Here $\nu$ is the $(j, j^{mw}, f, t)$-dependent effective nonlinear mobility of the vortex under the influence of the Lorentz force and it is expressed in terms of matrix continued fractions \cite{Shk14pcm}. In Eq. \eqref{eCVC}, the electric field is scaled to the electric field at the first Shapiro step, while the dc density $j$ and the mw density amplitude $j^{mw}$ to the depinning current density, frequency $f$ to the depinning frequency, coordinate $x$ to the WPP period, and temperature $t$ to the WPP depth.

We use the mean-square parameter $j_d=\sqrt{j_d^+ j_d^-}$ for the presentation of the data in dimensionless form allowing for a direct comparison of experiment with theory in Fig. \ref{fSteps}. The voltage responses are calculated by Eq. \eqref{eCVC} for a series of values of the asymmetry parameter $e$ at the reduced temperature $t = T/U_p = 0.002$ which corresponds to the pinning activation energy $U_p$ of about $5000$\,K \cite{Dob12njp} and the temperature $T = 0.98 T_c$ at which the experiment is conducted.

We now proceed to a deeper analysis of the mode-locking fringes reported in Fig. \ref{fLocking}(b). The appearance of these voltage peaks (for the positive dc polarity) and dips (for the negative dc polarity) can be explained by higher-order ratchet effects (labeled with ``1'' to ``4'' in the plot) as follows: We first consider the curve $V/V_0(j^{mw}/j^{mw}_d)$ in Fig. \ref{fLocking}(b) at $j/j_d = 0.05$, that is in the limit of very small dc biases. In the conventional ratchet effect resulting in the appearance of the ``1'' voltage peak, at a given ac amplitude $j^{mw} \gtrsim j^{mw}_{d~\mathrm{gentle}}$, a vortex overcomes the gentle slope of the WPP during one half of the ac period. However, the amplitude $j^{mw}$ is yet smaller than the strong-slope depinning current density and the vortex can not return into its original WPP well. This results in a net motion of the vortex by one WPP period and the associated rectified voltage. As the ac amplitude increases, the vortex can overcome the steep barrier of the WPP and the net motion disappears thus resulting in an almost full suppression of the rectified voltage. With a further increase of the ac amplitude the vortex can consequentially overcome two gentle barriers of WPP, and two options appear for its backward motion depending on the strength of the WPP asymmetry and the ac frequency, please refer to Fig. 2 of Supplementary: (i) The vortex can only overcome one barrier in the backward direction, and after one ac period it appears in the WPP well in \emph{one period} away from the original WPP well. (ii) The vortex can not overcome the barrier in the backward direction and it remains in the WPP well in \emph{two periods} away from the original WPP well. Obviously, with a further increase of the ac amplitude the ``splitting'' of ways increases for the vortex in what WPP well it will appear after one ac period  and this leads to a smearing of the higher-order ratchet peaks in the dc voltage.

The curve at $j = 0.25$ in Fig. \ref{fLocking}(b) corresponds to the particular interesting case of effective compensation of the intrinsic WPP asymmetry due to the difference in the steepness of the groove slopes by the external asymmetry of the potential induced by the dc bias. In consequence of this effective ``compensation of the anisotropies'' the ``floor'' of the rectified voltage can be adjusted to almost zero which is why the loading capability of the ratchet amounts to about $j/j_d = 0.25$ for sample A2. In all, the simulation results allow us not only to explain the main features observed in the experiment, but we find a very good quantitative agreement between the experimental data and theoretical modeling.

In conclusion, we have presented an approach allowing for the determination of a periodic pinning potential from the microwave power absorbed by vortices under dc bias reversal in superconductors with periodic pinning at low temperatures. For thicker films, the deduced coordinate dependences of the washboard pinning potentials $U(x)$ largely mimic the coordinate dependences of the film thickness, as visible from cross-sectional AFM line scans. The presented procedure allowed us to directly determine the asymmetry of the pinning potential and to further use it for modeling of voltage responses at $T = 0.98T_c$. Our findings pave a new route to the non-destructive evaluation of periodic pinning in superconductor thin films. Moreover, the reported approach should also apply to a broad class of systems whose evolution in time can be described by the coherent motion of (quasi)particles in a periodic potential.

\section{Methods}

\subsection{Film growth and characterization}

The $150\times500\,\mu$m$^2$ microstrips were fabricated by photolithography and Ar etching from epitaxial (110) Nb films on a-cut sapphire substrates. The films were grown by dc magnetron sputtering in a setup with a base pressure in the $10^{-8}$\,mbar range. In the sputtering process the substrate temperature was $850^\circ$C, the Ar pressure $4\times10^{-3}$\,mbar, and the growth rate was about $1$\,nm/s. X-ray diffraction measurements revealed the (110) orientation of the films \cite{Dob12tsf}. The epitaxy of the films has been confirmed by reflection high-energy electron diffraction. The as-grown films have a smooth surface with an rms surface roughness of less than $0.5$\,nm, as deduced from AFM scans in the range $1\,\mu$m$\times1\,\mu$m. The background pinning in the epitaxial (110) Nb films is very weak as compared to the anisotropic pinning induced by nanogrooves milled by FIB.  The parameters of the samples are in Table \ref{Table}.
\begin{table}[h!]
    \centering
    \begin{tabular}{|l|l|l|l|}
    \hline
    Parameter                             & Sample S & Sample A1 & Sample A2  \\
    \hline
    d, nm                                 & 40 & 56 & 70\\
    \hline
    $T_c$, K                              & 8.66 & 8.85 & 8.94 \\
    \hline
    $f_0$, GHz                            & 5.72 & 3.95 & 3.12\\
    \hline
    $j^+_d$ at $T = 0.3T_c$, MA/cm$^2$    & 0.75 & 0.7 & 0.52 \\
    \hline
    $j^-_d$ at $T = 0.3T_c$, MA/cm$^2$    & 0.75 & 0.91 & 1.25 \\
    \hline
    $j^-_d/j^+_d$, deduced from the CVC   & 1    & 1.3 & 2.4 \\
    \hline
    $e$, deduced asymmetry parameter      & 0    & 0.13 & 0.56 \\
    \hline
    \end{tabular}
    \caption{\label{Table} Main parameters of the investigated Nb microstrips.}
\end{table}

\subsection{Fabrication of nanogrooves}
Patterning of the samples was done in a high-resolution dual-beam scanning electron microscope (FEI, Nova Nanolab 600). In the patterning process, the asymmetry of the groove slopes was achieved by defining the grooves in the FIB bitmap file for sample S as a single line for the beam to pass, whereas a step-wise increasing number of FIB beam passes was assigned to each groove defined as a 3-step and a 5-step ``stair'' for samples A1 and A2, respectively. Due to blurring effects, the symmetric grooves in Sample S have rounded corners while smoothed straight slopes resulted instead of the ``stairs'' in samples A1 and A2. For all samples the beam parameters were $30$\,kV/$50$\,pA, $1$\,$\mu$s dwell time and $50$\,nm pitch. The grooves are parallel to the microstrip edges with an error of less than $0.2^\circ$. The microstrip width is an integer multiple number ($N=300$) of the nanopattern period to prevent possible ratchet effects due to the edge barrier asymmetry \cite{Pry06apl}.
\subsection{Atomic force microscopy}
A Nanosurf easyScan 2 atomic force microscope (AFM) under ambient conditions in non-contact, dynamic force mode was used. The cantilever tip was shaped like a polygon-based pyramid, with a tip radius of less than $7$\,nm (Nanosensors PPP-NCLR). Convolution effects due to the finite tip radius can be neglected, as is corroborated by the invariance of the AFM images taken with the cantilever scanning at different angles with respect to the grooves. The cross-sectional AFM profiles are the results of averaging of $250$\,line scans acquired for a scanning field of $500\times500$\,nm$^2$.

\subsection{Microwave spectroscopy}
Combined broadband mw and dc electrical measurements were done in a $^4$He cryostat with magnetic field $H$ directed perpendicular to the film surface. A custom-made cryogenic sample probe with coaxial cables was employed \cite{Dob15mst}. The mw signal was generated and analyzed by an Agilent E5071C vector network analyzer (VNA). The mw and dc signals were superimposed and uncoupled by using two bias-tees mounted at the VNA ports. The VNA operated in the frequency sweep mode, with 1548 frequency points scanned with an exponentially growing increment between $300$\,KHz and $14$\,GHz. For all frequencies $f$, the mw excitation power at the sample was $P = -20$\,dBm (10\,$\mu$W) kept by the VNA in accordance with the pre-saved calibration data for $S_{21ref}(f,T)$. At small fields $S_{21ref}$ does not depend on the magnetic field value $H$.

\subsection{Modeling of voltage responses}
The infinite matrix continued fractions for the calculation of the voltage responses by Eq. (29) of Ref. \cite{Shk14pcm} were approximated by matrix continued fractions of finite order. This has been done by putting $\mathbf{Q}_m = \mathbf{0}$ at some $m = M$, whereas the dimension of the submatrices $\mathbf{Q}_m$ and the vectors $\mathbf{C}_m$ was confined to some finite number $K$. Both $M$ and $K$ depend on temperature and mw density and on the number of harmonics to be taken into account. These numbers were chosen as $K = 51$ and $M=500$ for the reliable calculation of the components $F^1_k(\omega)$ for up to 10 harmonics, for $j^{mw}$ up to $5$, and for $t\equiv T/U_0 = 0.002$, respectively. This ensures a calculation accuracy of at least three digits. The Hall effect was neglected in the calculations.

\section{Acknowledgements}

Roland Sachser is acknowledged for automating the data acquisition and his help with nanopatterning. This work was supported through the DFG project DO1511/3-1 and conducted within the framework of the COST Action CA16218 (NANOCOHYBRI) of the European Cooperation in Science and Technology. Further, funding from the European Commission in the framework of the program Marie Sklodowska-Curie Actions -- Research and Innovation Staff Exchange (MSCA-RISE) under Grant Agreement No. 644348 (MagIC) is acknowledged.

\section{Author Contributions}

O.V.D. conducted the experiment and wrote the manuscript. O.V.D. and M.H. designated the samples. V.A.S. suggested the concept of the experiment and analyzed the data. R.V.V. took part in the discussion of the results and analyzed the electrical voltage responses. M.H., V.A.S., and R.V.V. participated in the manuscript writing.

\vspace*{3mm}
\section{Supplementary}
\subsection{Frequency dependence of the power absorption}
\label{SupplPower}
\begin{figure}[h!]
    \centering
    \includegraphics[width=1\linewidth]{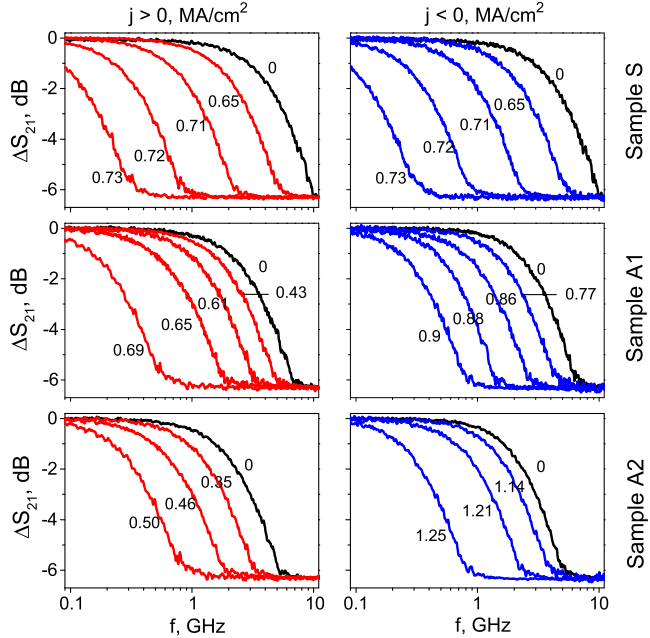}
    \caption{Frequency dependence of the relative change of the absolute value of the forward transmission coefficient $\Delta S_{21}(f)$ of all samples at positive (left column, red curves) and negative (right column, blue curves) dc densities, as indicated, at $T = 0.3T_c$ and the fundamental matching field $H_m = 7.2$\,mT. At dc biases of the positive polarity, the dc Lorentz force is directed against the gentle groove slope.}
    \label{fSupplSvF}
\end{figure}

\subsection{Mechanistic scenario for mode-locking fringes to appear}
\begin{figure}[h!]
    \centering
    \includegraphics[width=1\linewidth]{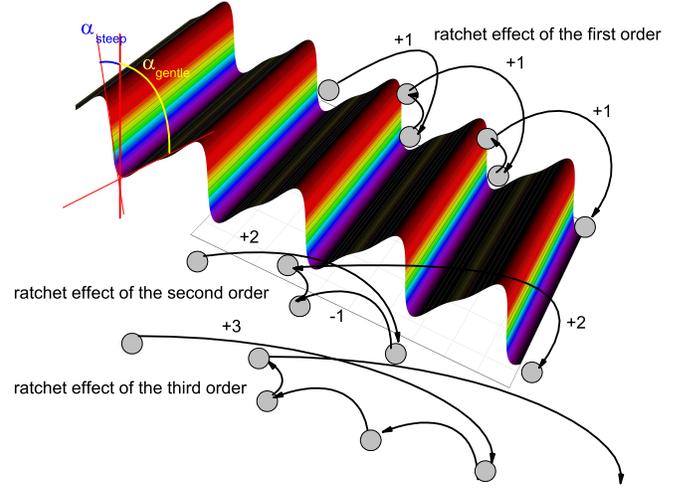}
    \caption{In the ratchet effect of the first order, during one ac period a vortex (grey circle) overcomes the right barrier during one half ac period, but the ac amplitude is not enough to let the vortex overcome the left barrier during the other half of the ac period. With increasing ac amplitude the vortex can overcome two barriers to right and either one (shown in the sketch) or no (not shown in the sketch) barrier to the left and so on. For the ratchet effect of the third order, only a vortex overcoming three barriers to the right and no barrier to the left is shown. Each order of the ratchet effect corresponds to a peak or dip (depending on the polarity of the applied dc bias) in the dependence of the dc voltage on the ac amplitude. The angles $\alpha_{\mathrm{steep}}$ and $\alpha_{\mathrm{gentle}}$ indicate the angles under which the slopes of the asymmetric WPP are tilted with respect to the vertical axis which stands for the pinning potential $U(x)$. For the orientation of the steep and gentle slopes shown in the sketch, the condition of the effective symmetrization $\alpha_{\mathrm{steep}} = \alpha_{\mathrm{gentle}}$ will be realized at some dc bias value resulting in a tilt of the pinning potential to the left. In reality, however, the assumption that the slopes of the potential can be modelled as planes is too crude. But here the angles $\alpha$ are introduced just for an obvious explanation how the internal asymmetry of the ratchet WPP can be ``compensated'' by the tilt induced by the dc bias. In the linear approximation, the physical meaning of these angles is the pinning forces for the respective groove slopes.}
    \label{fSupplSvF}
\end{figure}

\subsection{Determination of the pinning potential}
\label{SupplPotential}
The procedure of the determination of the coordinate dependence of the pinning potential is based on the Gittleman-Rosenblum (GR) model \cite{Git66prl} developed for zero temperature and generalized for the case of arbitrary dc and ac bias values \cite{Shk12inb}. In this model, the equation of motion for a vortex moving with velocity $v(t)$ in some pinning potential under the action of superimposed dc and high-frequency ac currents reads
\begin{equation}
    \label{eLE}
    \eta v(t) = F(t) + F_p,
\end{equation}
where $\eta$ is the vortex viscosity, $v(t)$ is the vortex velocity, and $F(t) = \frac{\Phi_0}{c} (j + j^{mw}(t))$ is the Lorentz force, where $j$ is the dc current density and $j^{mw}(t) \equiv j^{mw}\exp2\pi i f t$ with $j^{mw}$ is the amplitude of the microwave current. In Eq.~\eqref{eLE}, $F_p = -dU(x)/dx$ is the pinning force and $U(x)$ is the sought-for pinning potential with the depth $U_0$ and period $a$. Henceforth we scale $j$ to $j_d$, $U$ to $U_0$ and $x$ to $a$. Accordingly, the pinning potential is related \cite{Shk08mmt,Shk12inb} to the dc current-induced reduction of the depinning frequency via
\begin{equation}
    \label{eUx}
    U(x) = \int_0^x d x_0 \frac{j(x_0)}{j_d},
\end{equation}
where $x_0$ is the rest point of the vortex in the tilted pinning potential in the absence of the mw current. Further, the function $j(x_0)$ is the inverse function to $x_0(j)$ given by
\begin{equation}
    \label{eX0}
    x_0(j) = \int_0^{j} \frac {dj^\prime}{f_d(j^{\prime})/f_0},
\end{equation}
where $f_d(j^\prime)/f_0$ should be deduced from the experimental data.

We approximated the reduction of the depinning frequency for both dc polarities for sample S by the following expression
\begin{equation}
    \label{eDFvJS}
    f_d/f_0 = [1 - (j/j_d)^2]^{1/2},\qquad |j|< |j_d|.\\
\end{equation}
For the gentle-slope direction of the asymmetric potential of sample A1 probed by the positive halfwave of the ac current, the fit reads
\begin{equation}
    \label{eDFvJAsteep}
     f_d/f_0 = [1 - (j/j_d)^{3/2}]^{1/2},\quad 0 < j< j^+_d,
\end{equation}
while for its steep-slope direction probed by the positive halfwave of the ac current, the dependence reads
\begin{equation}
    \label{eDFvJAsteep}
     f_d/f_0 = [1 - (j/j_d)^3]^{1/2},\quad-j^-_d  < -j< 0.
\end{equation}

For the gentle-slope direction of the asymmetric potential of sample A2 probed by the positive halfwave of the ac current, the reduction of the depinning frequency can be fitted to
\begin{equation}
    \label{eDFvJAsteep}
     f_d/f_0 = [1 - (j/j_d)^{3/2}]^{2/3},\quad 0 < j< j^+_d,
\end{equation}
while for its steep-slope direction probed by the positive halfwave of the ac current, the approximation reads
\begin{equation}
    \label{eDFvJAsteep}
     f_d/f_0 = [1 - (j/j_d)^4]^{1/4},\quad-j^-_d  < -j< 0.
\end{equation}

The deduced dependence $U(x)/U_0$ in sample S is symmetric with respect to the line $x/a = 0.5$ and it fits very well to
\begin{equation}
    \label{eCosine}
     U(x)/U_0 = [1 - \cos2\pi x]/2.
\end{equation}
The deduced dependences $U(x)/U_0$ for samples A1 and A2 are asymmetric with respect to the line $x/a = 0.5$. For sample A1 $U(x)/U_0$ exhibits a maximum at $x\approx0.44$ and it can be satisfactory fitted to
\begin{equation}
    \label{eWeakRatchet}
    U(x)/U_0 = [(1 - \cos2\pi x) + 0.13(1-\sin4\pi x)/2]/2.
\end{equation}
The deduced dependence $U(x)/U_0$ for sample A2 is most strongly asymmetric. The curve $U(x)/U_0$ has a maximum at $x\approx0.32$ and it can be fitted rather well to
\begin{equation}
    \label{eStrongRatchet}
    U(x)/U_0 = [(1 - \cos2\pi x) + 0.5(1-\sin4\pi x)/2]/2.
\end{equation}
while the AFM profile suggests that also it can be fitted satisfactory to the same expression.


\begin{thebibliography}{54}%
\makeatletter
\providecommand \@ifxundefined [1]{%
 \@ifx{#1\undefined}
}%
\providecommand \@ifnum [1]{%
 \ifnum #1\expandafter \@firstoftwo
 \else \expandafter \@secondoftwo
 \fi
}%
\providecommand \@ifx [1]{%
 \ifx #1\expandafter \@firstoftwo
 \else \expandafter \@secondoftwo
 \fi
}%
\providecommand \natexlab [1]{#1}%
\providecommand \enquote  [1]{``#1''}%
\providecommand \bibnamefont  [1]{#1}%
\providecommand \bibfnamefont [1]{#1}%
\providecommand \citenamefont [1]{#1}%
\providecommand \href@noop [0]{\@secondoftwo}%
\providecommand \href [0]{\begingroup \@sanitize@url \@href}%
\providecommand \@href[1]{\@@startlink{#1}\@@href}%
\providecommand \@@href[1]{\endgroup#1\@@endlink}%
\providecommand \@sanitize@url [0]{\catcode `\\12\catcode `\$12\catcode
  `\&12\catcode `\#12\catcode `\^12\catcode `\_12\catcode `\%12\relax}%
\providecommand \@@startlink[1]{}%
\providecommand \@@endlink[0]{}%
\providecommand \url  [0]{\begingroup\@sanitize@url \@url }%
\providecommand \@url [1]{\endgroup\@href {#1}{\urlprefix }}%
\providecommand \urlprefix  [0]{URL }%
\providecommand \Eprint [0]{\href }%
\providecommand \doibase [0]{http://dx.doi.org/}%
\providecommand \selectlanguage [0]{\@gobble}%
\providecommand \bibinfo  [0]{\@secondoftwo}%
\providecommand \bibfield  [0]{\@secondoftwo}%
\providecommand \translation [1]{[#1]}%
\providecommand \BibitemOpen [0]{}%
\providecommand \bibitemStop [0]{}%
\providecommand \bibitemNoStop [0]{.\EOS\space}%
\providecommand \EOS [0]{\spacefactor3000\relax}%
\providecommand \BibitemShut  [1]{\csname bibitem#1\endcsname}%
\let\auto@bib@innerbib\@empty
\bibitem [{\citenamefont {Barone}\ and\ \citenamefont
  {Patterno}(1982)}]{Bar82boo}%
  \BibitemOpen
  \bibfield  {author} {\bibinfo {author} {\bibfnamefont {A.}~\bibnamefont
  {Barone}}\ and\ \bibinfo {author} {\bibfnamefont {G.}~\bibnamefont
  {Patterno}},\ }\href@noop {} {\emph {\bibinfo {title} {Physics and
  Applications of the Josephson Effect}}}\ (\bibinfo  {publisher} {John Wiley
  \& Sons, New York},\ \bibinfo {year} {1982})\BibitemShut {NoStop}%
\bibitem [{\citenamefont {Fulde}\ \emph {et~al.}(1975)\citenamefont {Fulde},
  \citenamefont {Pietronero}, \citenamefont {Schneider},\ and\ \citenamefont
  {Str\"assler}}]{Ful75prl}%
  \BibitemOpen
  \bibfield  {author} {\bibinfo {author} {\bibfnamefont {P.}~\bibnamefont
  {Fulde}}, \bibinfo {author} {\bibfnamefont {L.}~\bibnamefont {Pietronero}},
  \bibinfo {author} {\bibfnamefont {W.~R.}\ \bibnamefont {Schneider}}, \ and\
  \bibinfo {author} {\bibfnamefont {S.}~\bibnamefont {Str\"assler}},\ }\href
  {\doibase 10.1103/PhysRevLett.35.1776} {\bibfield  {journal} {\bibinfo
  {journal} {Phys. Rev. Lett.}\ }\textbf {\bibinfo {volume} {35}},\ \bibinfo
  {pages} {1776} (\bibinfo {year} {1975})}\BibitemShut {NoStop}%
\bibitem [{\citenamefont {Chow}\ \emph {et~al.}(1985)\citenamefont {Chow},
  \citenamefont {Gea-Banacloche}, \citenamefont {Pedrotti}, \citenamefont
  {Sanders}, \citenamefont {Schleich},\ and\ \citenamefont
  {Scully}}]{Cho85rmp}%
  \BibitemOpen
  \bibfield  {author} {\bibinfo {author} {\bibfnamefont {W.~W.}\ \bibnamefont
  {Chow}}, \bibinfo {author} {\bibfnamefont {J.}~\bibnamefont
  {Gea-Banacloche}}, \bibinfo {author} {\bibfnamefont {L.~M.}\ \bibnamefont
  {Pedrotti}}, \bibinfo {author} {\bibfnamefont {V.~E.}\ \bibnamefont
  {Sanders}}, \bibinfo {author} {\bibfnamefont {W.}~\bibnamefont {Schleich}}, \
  and\ \bibinfo {author} {\bibfnamefont {M.~O.}\ \bibnamefont {Scully}},\
  }\href {\doibase 10.1103/RevModPhys.57.61} {\bibfield  {journal} {\bibinfo
  {journal} {Rev. Mod. Phys.}\ }\textbf {\bibinfo {volume} {57}},\ \bibinfo
  {pages} {61} (\bibinfo {year} {1985})}\BibitemShut {NoStop}%
\bibitem [{\citenamefont {Barthel}\ \emph {et~al.}(1993)\citenamefont
  {Barthel}, \citenamefont {Kriza}, \citenamefont {Quirion}, \citenamefont
  {Wzietek}, \citenamefont {J\'erome}, \citenamefont {Christensen},
  \citenamefont {J\o{}rgensen},\ and\ \citenamefont {Bechgaard}}]{Bar93prl}%
  \BibitemOpen
  \bibfield  {author} {\bibinfo {author} {\bibfnamefont {E.}~\bibnamefont
  {Barthel}}, \bibinfo {author} {\bibfnamefont {G.}~\bibnamefont {Kriza}},
  \bibinfo {author} {\bibfnamefont {G.}~\bibnamefont {Quirion}}, \bibinfo
  {author} {\bibfnamefont {P.}~\bibnamefont {Wzietek}}, \bibinfo {author}
  {\bibfnamefont {D.}~\bibnamefont {J\'erome}}, \bibinfo {author}
  {\bibfnamefont {J.~B.}\ \bibnamefont {Christensen}}, \bibinfo {author}
  {\bibfnamefont {M.}~\bibnamefont {J\o{}rgensen}}, \ and\ \bibinfo {author}
  {\bibfnamefont {K.}~\bibnamefont {Bechgaard}},\ }\href {\doibase
  10.1103/PhysRevLett.71.2825} {\bibfield  {journal} {\bibinfo  {journal}
  {Phys. Rev. Lett.}\ }\textbf {\bibinfo {volume} {71}},\ \bibinfo {pages}
  {2825} (\bibinfo {year} {1993})}\BibitemShut {NoStop}%
\bibitem [{\citenamefont {Zybtsev}\ and\ \citenamefont
  {Pokrovskii}(2013)}]{Zyb13prb}%
  \BibitemOpen
  \bibfield  {author} {\bibinfo {author} {\bibfnamefont {S.~G.}\ \bibnamefont
  {Zybtsev}}\ and\ \bibinfo {author} {\bibfnamefont {V.~Y.}\ \bibnamefont
  {Pokrovskii}},\ }\href {\doibase 10.1103/PhysRevB.88.125144} {\bibfield
  {journal} {\bibinfo  {journal} {Phys. Rev. B}\ }\textbf {\bibinfo {volume}
  {88}},\ \bibinfo {pages} {125144} (\bibinfo {year} {2013})}\BibitemShut
  {NoStop}%
\bibitem [{\citenamefont {Risken}(1989)}]{Ris89boo}%
  \BibitemOpen
  \bibfield  {author} {\bibinfo {author} {\bibfnamefont {H.}~\bibnamefont
  {Risken}},\ }\href@noop {} {\emph {\bibinfo {title} {{The Fokker-Planck
  Equation}}}}\ (\bibinfo  {publisher} {Springer, Berlin},\ \bibinfo {year}
  {1989})\BibitemShut {NoStop}%
\bibitem [{\citenamefont {P\'erez-Junquera}\ \emph {et~al.}(2008)\citenamefont
  {P\'erez-Junquera}, \citenamefont {Marconi}, \citenamefont {Kolton},
  \citenamefont {\'Alvarez-Prado}, \citenamefont {Souche}, \citenamefont
  {Alija}, \citenamefont {V\'elez}, \citenamefont {Anguita}, \citenamefont
  {Alameda}, \citenamefont {Mart\'in},\ and\ \citenamefont
  {Parrondo}}]{Per08prl}%
  \BibitemOpen
  \bibfield  {author} {\bibinfo {author} {\bibfnamefont {A.}~\bibnamefont
  {P\'erez-Junquera}}, \bibinfo {author} {\bibfnamefont {V.~I.}\ \bibnamefont
  {Marconi}}, \bibinfo {author} {\bibfnamefont {A.~B.}\ \bibnamefont {Kolton}},
  \bibinfo {author} {\bibfnamefont {L.~M.}\ \bibnamefont {\'Alvarez-Prado}},
  \bibinfo {author} {\bibfnamefont {Y.}~\bibnamefont {Souche}}, \bibinfo
  {author} {\bibfnamefont {A.}~\bibnamefont {Alija}}, \bibinfo {author}
  {\bibfnamefont {M.}~\bibnamefont {V\'elez}}, \bibinfo {author} {\bibfnamefont
  {J.~V.}\ \bibnamefont {Anguita}}, \bibinfo {author} {\bibfnamefont {J.~M.}\
  \bibnamefont {Alameda}}, \bibinfo {author} {\bibfnamefont {J.~I.}\
  \bibnamefont {Mart\'in}}, \ and\ \bibinfo {author} {\bibfnamefont {J.~M.~R.}\
  \bibnamefont {Parrondo}},\ }\href {\doibase 10.1103/PhysRevLett.100.037203}
  {\bibfield  {journal} {\bibinfo  {journal} {Phys. Rev. Lett.}\ }\textbf
  {\bibinfo {volume} {100}},\ \bibinfo {pages} {037203} (\bibinfo {year}
  {2008})}\BibitemShut {NoStop}%
\bibitem [{\citenamefont {Titov}\ \emph {et~al.}(2005)\citenamefont {Titov},
  \citenamefont {Kachkachi}, \citenamefont {Kalmykov},\ and\ \citenamefont
  {Coffey}}]{Tit05prb}%
  \BibitemOpen
  \bibfield  {author} {\bibinfo {author} {\bibfnamefont {S.~V.}\ \bibnamefont
  {Titov}}, \bibinfo {author} {\bibfnamefont {H.}~\bibnamefont {Kachkachi}},
  \bibinfo {author} {\bibfnamefont {Y.~P.}\ \bibnamefont {Kalmykov}}, \ and\
  \bibinfo {author} {\bibfnamefont {W.~T.}\ \bibnamefont {Coffey}},\ }\href
  {\doibase 10.1103/PhysRevB.72.134425} {\bibfield  {journal} {\bibinfo
  {journal} {Phys. Rev. B}\ }\textbf {\bibinfo {volume} {72}},\ \bibinfo
  {pages} {134425} (\bibinfo {year} {2005})}\BibitemShut {NoStop}%
\bibitem [{\citenamefont {Evstigneev}\ \emph {et~al.}(2008)\citenamefont
  {Evstigneev}, \citenamefont {Zvyagolskaya}, \citenamefont {Bleil},
  \citenamefont {Eichhorn}, \citenamefont {Bechinger},\ and\ \citenamefont
  {Reimann}}]{Evs08pre}%
  \BibitemOpen
  \bibfield  {author} {\bibinfo {author} {\bibfnamefont {M.}~\bibnamefont
  {Evstigneev}}, \bibinfo {author} {\bibfnamefont {O.}~\bibnamefont
  {Zvyagolskaya}}, \bibinfo {author} {\bibfnamefont {S.}~\bibnamefont {Bleil}},
  \bibinfo {author} {\bibfnamefont {R.}~\bibnamefont {Eichhorn}}, \bibinfo
  {author} {\bibfnamefont {C.}~\bibnamefont {Bechinger}}, \ and\ \bibinfo
  {author} {\bibfnamefont {P.}~\bibnamefont {Reimann}},\ }\href {\doibase
  10.1103/PhysRevE.77.041107} {\bibfield  {journal} {\bibinfo  {journal} {Phys.
  Rev. E}\ }\textbf {\bibinfo {volume} {77}},\ \bibinfo {pages} {041107}
  (\bibinfo {year} {2008})}\BibitemShut {NoStop}%
\bibitem [{\citenamefont {Brandt}(1995)}]{Bra95rpp}%
  \BibitemOpen
  \bibfield  {author} {\bibinfo {author} {\bibfnamefont {E.~H.}\ \bibnamefont
  {Brandt}},\ }\href {http://stacks.iop.org/0034-4885/58/i=11/a=003} {\bibfield
   {journal} {\bibinfo  {journal} {Rep. Progr. Phys.}\ }\textbf {\bibinfo
  {volume} {58}},\ \bibinfo {pages} {1465} (\bibinfo {year}
  {1995})}\BibitemShut {NoStop}%
\bibitem [{\citenamefont {Plourde}(2009)}]{Plo09tas}%
  \BibitemOpen
  \bibfield  {author} {\bibinfo {author} {\bibfnamefont {B.~L.~T.}\
  \bibnamefont {Plourde}},\ }\href {\doibase doi:10.1109/TASC.2009.2028873}
  {\bibfield  {journal} {\bibinfo  {journal} {IEEE Trans. Appl. Supercond.}\
  }\textbf {\bibinfo {volume} {19}},\ \bibinfo {pages} {3698} (\bibinfo {year}
  {2009})}\BibitemShut {NoStop}%
\bibitem [{\citenamefont {Cuadra-Solis}\ \emph {et~al.}(2014)\citenamefont
  {Cuadra-Solis}, \citenamefont {Garcia-Santiago}, \citenamefont {Hernandez},
  \citenamefont {Tejada}, \citenamefont {Vanacken},\ and\ \citenamefont
  {Moshchalkov}}]{Sol14prb}%
  \BibitemOpen
  \bibfield  {author} {\bibinfo {author} {\bibfnamefont {P.-d.-J.}\
  \bibnamefont {Cuadra-Solis}}, \bibinfo {author} {\bibfnamefont
  {A.}~\bibnamefont {Garcia-Santiago}}, \bibinfo {author} {\bibfnamefont
  {J.~M.}\ \bibnamefont {Hernandez}}, \bibinfo {author} {\bibfnamefont
  {J.}~\bibnamefont {Tejada}}, \bibinfo {author} {\bibfnamefont
  {J.}~\bibnamefont {Vanacken}}, \ and\ \bibinfo {author} {\bibfnamefont
  {V.~V.}\ \bibnamefont {Moshchalkov}},\ }\href {\doibase
  10.1103/PhysRevB.89.054517} {\bibfield  {journal} {\bibinfo  {journal} {Phys.
  Rev. B}\ }\textbf {\bibinfo {volume} {89}},\ \bibinfo {pages} {054517}
  (\bibinfo {year} {2014})}\BibitemShut {NoStop}%
\bibitem [{\citenamefont {Pryadun}\ \emph {et~al.}(2006)\citenamefont
  {Pryadun}, \citenamefont {Sierra}, \citenamefont {Aliev}, \citenamefont
  {Golubovic},\ and\ \citenamefont {Moshchalkov}}]{Pry06apl}%
  \BibitemOpen
  \bibfield  {author} {\bibinfo {author} {\bibfnamefont {V.~V.}\ \bibnamefont
  {Pryadun}}, \bibinfo {author} {\bibfnamefont {J.}~\bibnamefont {Sierra}},
  \bibinfo {author} {\bibfnamefont {F.~G.}\ \bibnamefont {Aliev}}, \bibinfo
  {author} {\bibfnamefont {D.~S.}\ \bibnamefont {Golubovic}}, \ and\ \bibinfo
  {author} {\bibfnamefont {V.~V.}\ \bibnamefont {Moshchalkov}},\ }\href
  {\doibase http://dx.doi.org/10.1063/1.2171788} {\bibfield  {journal}
  {\bibinfo  {journal} {Appl. Phys. Lett.}\ }\textbf {\bibinfo {volume} {88}},\
  \bibinfo {eid} {062517} (\bibinfo {year} {2006})}\BibitemShut {NoStop}%
\bibitem [{\citenamefont {Jelic}\ \emph {et~al.}(2015)\citenamefont {Jelic},
  \citenamefont {Milosevic}, \citenamefont {Van~de Vondel},\ and\ \citenamefont
  {Silhanek}}]{Jel15nsr}%
  \BibitemOpen
  \bibfield  {author} {\bibinfo {author} {\bibfnamefont {Z.~L.}\ \bibnamefont
  {Jelic}}, \bibinfo {author} {\bibfnamefont {M.~V.}\ \bibnamefont
  {Milosevic}}, \bibinfo {author} {\bibfnamefont {J.}~\bibnamefont {Van~de
  Vondel}}, \ and\ \bibinfo {author} {\bibfnamefont {A.~V.}\ \bibnamefont
  {Silhanek}},\ }\href {http://dx.doi.org/10.1038/srep14604} {\bibfield
  {journal} {\bibinfo  {journal} {Sci. Rep.}\ }\textbf {\bibinfo {volume}
  {5}},\ \bibinfo {pages} {14604 EP } (\bibinfo {year} {2015})},\ \bibinfo
  {note} {article}\BibitemShut {NoStop}%
\bibitem [{\citenamefont {Lara}\ \emph {et~al.}(2015)\citenamefont {Lara},
  \citenamefont {Aliev}, \citenamefont {Silhanek},\ and\ \citenamefont
  {Moshchalkov}}]{Lar15nsr}%
  \BibitemOpen
  \bibfield  {author} {\bibinfo {author} {\bibfnamefont {A.}~\bibnamefont
  {Lara}}, \bibinfo {author} {\bibfnamefont {F.~G.}\ \bibnamefont {Aliev}},
  \bibinfo {author} {\bibfnamefont {A.~V.}\ \bibnamefont {Silhanek}}, \ and\
  \bibinfo {author} {\bibfnamefont {V.~V.}\ \bibnamefont {Moshchalkov}},\
  }\href {\doibase http://dx.doi.org/10.1038/srep09187} {\bibfield  {journal}
  {\bibinfo  {journal} {Sci. Rep.}\ }\textbf {\bibinfo {volume} {5}},\ \bibinfo
  {pages} {9187} (\bibinfo {year} {2015})}\BibitemShut {NoStop}%
\bibitem [{\citenamefont {Silhanek}\ \emph {et~al.}(2012)\citenamefont
  {Silhanek}, \citenamefont {Leo}, \citenamefont {Grimaldi}, \citenamefont
  {Berdiyorov}, \citenamefont {Milosevic}, \citenamefont {Nigro}, \citenamefont
  {Pace}, \citenamefont {Verellen}, \citenamefont {Gillijns}, \citenamefont
  {Metlushko}, \citenamefont {Ili}, \citenamefont {Zhu},\ and\ \citenamefont
  {Moshchalkov}}]{Sil12njp}%
  \BibitemOpen
  \bibfield  {author} {\bibinfo {author} {\bibfnamefont {A.~V.}\ \bibnamefont
  {Silhanek}}, \bibinfo {author} {\bibfnamefont {A.}~\bibnamefont {Leo}},
  \bibinfo {author} {\bibfnamefont {G.}~\bibnamefont {Grimaldi}}, \bibinfo
  {author} {\bibfnamefont {G.~R.}\ \bibnamefont {Berdiyorov}}, \bibinfo
  {author} {\bibfnamefont {M.~V.}\ \bibnamefont {Milosevic}}, \bibinfo {author}
  {\bibfnamefont {A.}~\bibnamefont {Nigro}}, \bibinfo {author} {\bibfnamefont
  {S.}~\bibnamefont {Pace}}, \bibinfo {author} {\bibfnamefont {N.}~\bibnamefont
  {Verellen}}, \bibinfo {author} {\bibfnamefont {W.}~\bibnamefont {Gillijns}},
  \bibinfo {author} {\bibfnamefont {V.}~\bibnamefont {Metlushko}}, \bibinfo
  {author} {\bibfnamefont {B.}~\bibnamefont {Ili}}, \bibinfo {author}
  {\bibfnamefont {X.}~\bibnamefont {Zhu}}, \ and\ \bibinfo {author}
  {\bibfnamefont {V.~V.}\ \bibnamefont {Moshchalkov}},\ }\href
  {http://stacks.iop.org/1367-2630/14/i=5/a=053006} {\bibfield  {journal}
  {\bibinfo  {journal} {New J. Phys.}\ }\textbf {\bibinfo {volume} {14}},\
  \bibinfo {pages} {053006} (\bibinfo {year} {2012})}\BibitemShut {NoStop}%
\bibitem [{\citenamefont {Grimaldi}\ \emph {et~al.}(2015)\citenamefont
  {Grimaldi}, \citenamefont {Leo}, \citenamefont {Sabatino}, \citenamefont
  {Carapella}, \citenamefont {Nigro}, \citenamefont {Pace}, \citenamefont
  {Moshchalkov},\ and\ \citenamefont {Silhanek}}]{Gri15prb}%
  \BibitemOpen
  \bibfield  {author} {\bibinfo {author} {\bibfnamefont {G.}~\bibnamefont
  {Grimaldi}}, \bibinfo {author} {\bibfnamefont {A.}~\bibnamefont {Leo}},
  \bibinfo {author} {\bibfnamefont {P.}~\bibnamefont {Sabatino}}, \bibinfo
  {author} {\bibfnamefont {G.}~\bibnamefont {Carapella}}, \bibinfo {author}
  {\bibfnamefont {A.}~\bibnamefont {Nigro}}, \bibinfo {author} {\bibfnamefont
  {S.}~\bibnamefont {Pace}}, \bibinfo {author} {\bibfnamefont {V.~V.}\
  \bibnamefont {Moshchalkov}}, \ and\ \bibinfo {author} {\bibfnamefont {A.~V.}\
  \bibnamefont {Silhanek}},\ }\href {\doibase 10.1103/PhysRevB.92.024513}
  {\bibfield  {journal} {\bibinfo  {journal} {Phys. Rev. B}\ }\textbf {\bibinfo
  {volume} {92}},\ \bibinfo {pages} {024513} (\bibinfo {year}
  {2015})}\BibitemShut {NoStop}%
\bibitem [{\citenamefont {Shklovskij}\ \emph {et~al.}(2017)\citenamefont
  {Shklovskij}, \citenamefont {Nazipova},\ and\ \citenamefont
  {Dobrovolskiy}}]{Shk17prb}%
  \BibitemOpen
  \bibfield  {author} {\bibinfo {author} {\bibfnamefont {V.~A.}\ \bibnamefont
  {Shklovskij}}, \bibinfo {author} {\bibfnamefont {A.~P.}\ \bibnamefont
  {Nazipova}}, \ and\ \bibinfo {author} {\bibfnamefont {O.~V.}\ \bibnamefont
  {Dobrovolskiy}},\ }\href {\doibase 10.1103/PhysRevB.95.184517} {\bibfield
  {journal} {\bibinfo  {journal} {Phys. Rev. B}\ }\textbf {\bibinfo {volume}
  {95}},\ \bibinfo {pages} {184517} (\bibinfo {year} {2017})}\BibitemShut
  {NoStop}%
\bibitem [{\citenamefont {Dobrovolskiy}\ \emph {et~al.}(2017)\citenamefont
  {Dobrovolskiy}, \citenamefont {Shklovskij}, \citenamefont {Hanefeld},
  \citenamefont {Z\"orb}, \citenamefont {K\"ohs},\ and\ \citenamefont
  {Huth}}]{Dob17sst}%
  \BibitemOpen
  \bibfield  {author} {\bibinfo {author} {\bibfnamefont {O.~V.}\ \bibnamefont
  {Dobrovolskiy}}, \bibinfo {author} {\bibfnamefont {V.~A.}\ \bibnamefont
  {Shklovskij}}, \bibinfo {author} {\bibfnamefont {M.}~\bibnamefont
  {Hanefeld}}, \bibinfo {author} {\bibfnamefont {M.}~\bibnamefont {Z\"orb}},
  \bibinfo {author} {\bibfnamefont {L.}~\bibnamefont {K\"ohs}}, \ and\ \bibinfo
  {author} {\bibfnamefont {M.}~\bibnamefont {Huth}},\ }\href
  {http://stacks.iop.org/0953-2048/30/i=8/a=085002} {\bibfield  {journal}
  {\bibinfo  {journal} {Supercond. Sci. Technol.}\ }\textbf {\bibinfo {volume}
  {30}},\ \bibinfo {pages} {085002} (\bibinfo {year} {2017})}\BibitemShut
  {NoStop}%
\bibitem [{\citenamefont {Villegas}\ \emph {et~al.}(2003)\citenamefont
  {Villegas}, \citenamefont {Savel'ev}, \citenamefont {Nori}, \citenamefont
  {Gonzalez}, \citenamefont {Anguita}, \citenamefont {Garcia},\ and\
  \citenamefont {Vicent}}]{Vil03sci}%
  \BibitemOpen
  \bibfield  {author} {\bibinfo {author} {\bibfnamefont {J.~E.}\ \bibnamefont
  {Villegas}}, \bibinfo {author} {\bibfnamefont {S.}~\bibnamefont {Savel'ev}},
  \bibinfo {author} {\bibfnamefont {F.}~\bibnamefont {Nori}}, \bibinfo {author}
  {\bibfnamefont {E.~M.}\ \bibnamefont {Gonzalez}}, \bibinfo {author}
  {\bibfnamefont {J.~V.}\ \bibnamefont {Anguita}}, \bibinfo {author}
  {\bibfnamefont {R.}~\bibnamefont {Garcia}}, \ and\ \bibinfo {author}
  {\bibfnamefont {J.~L.}\ \bibnamefont {Vicent}},\ }\href {\doibase
  10.1126/science.1090390} {\bibfield  {journal} {\bibinfo  {journal}
  {Science}\ }\textbf {\bibinfo {volume} {302}},\ \bibinfo {pages} {1188}
  (\bibinfo {year} {2003})}\BibitemShut {NoStop}%
\bibitem [{\citenamefont {Vlasko-Vlasov}\ \emph {et~al.}(2016)\citenamefont
  {Vlasko-Vlasov}, \citenamefont {Colauto}, \citenamefont {Benseman},
  \citenamefont {Rosenmann},\ and\ \citenamefont {Kwok}}]{Vla16nsr}%
  \BibitemOpen
  \bibfield  {author} {\bibinfo {author} {\bibfnamefont {V.~K.}\ \bibnamefont
  {Vlasko-Vlasov}}, \bibinfo {author} {\bibfnamefont {F.}~\bibnamefont
  {Colauto}}, \bibinfo {author} {\bibfnamefont {T.}~\bibnamefont {Benseman}},
  \bibinfo {author} {\bibfnamefont {D.}~\bibnamefont {Rosenmann}}, \ and\
  \bibinfo {author} {\bibfnamefont {W.-K.}\ \bibnamefont {Kwok}},\ }\href
  {http://dx.doi.org/10.1038/srep36847} {\bibfield  {journal} {\bibinfo
  {journal} {Sci. Rep.}\ }\textbf {\bibinfo {volume} {6}},\ \bibinfo {pages}
  {36847 EP } (\bibinfo {year} {2016})},\ \bibinfo {note} {article}\BibitemShut
  {NoStop}%
\bibitem [{\citenamefont {Dobrovolskiy}\ and\ \citenamefont
  {Huth}(2015)}]{Dob15apl}%
  \BibitemOpen
  \bibfield  {author} {\bibinfo {author} {\bibfnamefont {O.~V.}\ \bibnamefont
  {Dobrovolskiy}}\ and\ \bibinfo {author} {\bibfnamefont {M.}~\bibnamefont
  {Huth}},\ }\href {\doibase http://dx.doi.org/10.1063/1.4917229} {\bibfield
  {journal} {\bibinfo  {journal} {Appl. Phys. Lett.}\ }\textbf {\bibinfo
  {volume} {106}},\ \bibinfo {pages} {142601} (\bibinfo {year}
  {2015})}\BibitemShut {NoStop}%
\bibitem [{\citenamefont {Gillijns}\ \emph {et~al.}(2007)\citenamefont
  {Gillijns}, \citenamefont {Silhanek}, \citenamefont {Moshchalkov},
  \citenamefont {Reichhardt},\ and\ \citenamefont {Reichhardt}}]{Gil07prl}%
  \BibitemOpen
  \bibfield  {author} {\bibinfo {author} {\bibfnamefont {W.}~\bibnamefont
  {Gillijns}}, \bibinfo {author} {\bibfnamefont {A.~V.}\ \bibnamefont
  {Silhanek}}, \bibinfo {author} {\bibfnamefont {V.~V.}\ \bibnamefont
  {Moshchalkov}}, \bibinfo {author} {\bibfnamefont {C.~J.~O.}\ \bibnamefont
  {Reichhardt}}, \ and\ \bibinfo {author} {\bibfnamefont {C.}~\bibnamefont
  {Reichhardt}},\ }\href {\doibase 10.1103/PhysRevLett.99.247002} {\bibfield
  {journal} {\bibinfo  {journal} {Phys. Rev. Lett.}\ }\textbf {\bibinfo
  {volume} {99}},\ \bibinfo {pages} {247002} (\bibinfo {year}
  {2007})}\BibitemShut {NoStop}%
\bibitem [{\citenamefont {Togawa}\ \emph {et~al.}(2005)\citenamefont {Togawa},
  \citenamefont {Harada}, \citenamefont {Akashi}, \citenamefont {Kasai},
  \citenamefont {Matsuda}, \citenamefont {Nori}, \citenamefont {Maeda},\ and\
  \citenamefont {Tonomura}}]{Tog05prl}%
  \BibitemOpen
  \bibfield  {author} {\bibinfo {author} {\bibfnamefont {Y.}~\bibnamefont
  {Togawa}}, \bibinfo {author} {\bibfnamefont {K.}~\bibnamefont {Harada}},
  \bibinfo {author} {\bibfnamefont {T.}~\bibnamefont {Akashi}}, \bibinfo
  {author} {\bibfnamefont {H.}~\bibnamefont {Kasai}}, \bibinfo {author}
  {\bibfnamefont {T.}~\bibnamefont {Matsuda}}, \bibinfo {author} {\bibfnamefont
  {F.}~\bibnamefont {Nori}}, \bibinfo {author} {\bibfnamefont {A.}~\bibnamefont
  {Maeda}}, \ and\ \bibinfo {author} {\bibfnamefont {A.}~\bibnamefont
  {Tonomura}},\ }\href {\doibase 10.1103/PhysRevLett.95.087002} {\bibfield
  {journal} {\bibinfo  {journal} {Phys. Rev. Lett.}\ }\textbf {\bibinfo
  {volume} {95}},\ \bibinfo {pages} {087002} (\bibinfo {year}
  {2005})}\BibitemShut {NoStop}%
\bibitem [{\citenamefont {Dobrovolskiy}(2017)}]{Dob17pcs}%
  \BibitemOpen
  \bibfield  {author} {\bibinfo {author} {\bibfnamefont {O.~V.}\ \bibnamefont
  {Dobrovolskiy}},\ }\href {\doibase
  http://doi.org/10.1016/j.physc.2016.07.008} {\bibfield  {journal} {\bibinfo
  {journal} {Physica C}\ }\textbf {\bibinfo {volume} {533}},\ \bibinfo {pages}
  {80} (\bibinfo {year} {2017})}\BibitemShut {NoStop}%
\bibitem [{\citenamefont {Dobrovolskiy}\ \emph
  {et~al.}(2015{\natexlab{a}})\citenamefont {Dobrovolskiy}, \citenamefont
  {Huth},\ and\ \citenamefont {Shklovskij}}]{Dob15met}%
  \BibitemOpen
  \bibfield  {author} {\bibinfo {author} {\bibfnamefont {O.~V.}\ \bibnamefont
  {Dobrovolskiy}}, \bibinfo {author} {\bibfnamefont {M.}~\bibnamefont {Huth}},
  \ and\ \bibinfo {author} {\bibfnamefont {V.~A.}\ \bibnamefont {Shklovskij}},\
  }\href {\doibase http://dx.doi.org/10.1063/1.4934487} {\bibfield  {journal}
  {\bibinfo  {journal} {Appl. Phys. Lett.}\ }\textbf {\bibinfo {volume}
  {107}},\ \bibinfo {pages} {162603} (\bibinfo {year}
  {2015}{\natexlab{a}})}\BibitemShut {NoStop}%
\bibitem [{\citenamefont {Dobrovolskiy}\ \emph {et~al.}(2010)\citenamefont
  {Dobrovolskiy}, \citenamefont {Huth},\ and\ \citenamefont
  {Shklovskij}}]{Dob10sst}%
  \BibitemOpen
  \bibfield  {author} {\bibinfo {author} {\bibfnamefont {O.~V.}\ \bibnamefont
  {Dobrovolskiy}}, \bibinfo {author} {\bibfnamefont {M.}~\bibnamefont {Huth}},
  \ and\ \bibinfo {author} {\bibfnamefont {V.~A.}\ \bibnamefont {Shklovskij}},\
  }\href {\doibase doi:10.1088/0953-2048/23/12/125014} {\bibfield  {journal}
  {\bibinfo  {journal} {Supercond. Sci. Technol.}\ }\textbf {\bibinfo {volume}
  {23}},\ \bibinfo {pages} {125014} (\bibinfo {year} {2010})}\BibitemShut
  {NoStop}%
\bibitem [{\citenamefont {Campbell}\ and\ \citenamefont
  {Evetts}(1972)}]{Cam72aip}%
  \BibitemOpen
  \bibfield  {author} {\bibinfo {author} {\bibfnamefont {A.~M.}\ \bibnamefont
  {Campbell}}\ and\ \bibinfo {author} {\bibfnamefont {J.~E.}\ \bibnamefont
  {Evetts}},\ }\href {\doibase 10.1080/00018737200101288} {\bibfield  {journal}
  {\bibinfo  {journal} {Adv. Phys.}\ }\textbf {\bibinfo {volume} {21}},\
  \bibinfo {pages} {199} (\bibinfo {year} {1972})}\BibitemShut {NoStop}%
\bibitem [{\citenamefont {Lowell}(1972)}]{Low72jpf}%
  \BibitemOpen
  \bibfield  {author} {\bibinfo {author} {\bibfnamefont {J.}~\bibnamefont
  {Lowell}},\ }\href {http://stacks.iop.org/0305-4608/2/i=3/a=023} {\bibfield
  {journal} {\bibinfo  {journal} {J. Phys. F: Metal Phys.}\ }\textbf {\bibinfo
  {volume} {2}},\ \bibinfo {pages} {559} (\bibinfo {year} {1972})}\BibitemShut
  {NoStop}%
\bibitem [{\citenamefont {Auslaender}\ \emph {et~al.}(2009)\citenamefont
  {Auslaender}, \citenamefont {Luan}, \citenamefont {Straver}, \citenamefont
  {Hoffman}, \citenamefont {Koshnick}, \citenamefont {Zeldov}, \citenamefont
  {Bonn}, \citenamefont {Liang}, \citenamefont {Hardy},\ and\ \citenamefont
  {Moler}}]{Aus09nph}%
  \BibitemOpen
  \bibfield  {author} {\bibinfo {author} {\bibfnamefont {O.~M.}\ \bibnamefont
  {Auslaender}}, \bibinfo {author} {\bibfnamefont {L.}~\bibnamefont {Luan}},
  \bibinfo {author} {\bibfnamefont {E.~W.~J.}\ \bibnamefont {Straver}},
  \bibinfo {author} {\bibfnamefont {J.~E.}\ \bibnamefont {Hoffman}}, \bibinfo
  {author} {\bibfnamefont {N.~C.}\ \bibnamefont {Koshnick}}, \bibinfo {author}
  {\bibfnamefont {E.}~\bibnamefont {Zeldov}}, \bibinfo {author} {\bibfnamefont
  {D.~A.}\ \bibnamefont {Bonn}}, \bibinfo {author} {\bibfnamefont
  {R.}~\bibnamefont {Liang}}, \bibinfo {author} {\bibfnamefont {W.~N.}\
  \bibnamefont {Hardy}}, \ and\ \bibinfo {author} {\bibfnamefont {K.~A.}\
  \bibnamefont {Moler}},\ }\href {\doibase 10.1038/nphys1127} {\bibfield
  {journal} {\bibinfo  {journal} {Nat Phys}\ }\textbf {\bibinfo {volume} {5}},\
  \bibinfo {pages} {35} (\bibinfo {year} {2009})}\BibitemShut {NoStop}%
\bibitem [{\citenamefont {Embon}\ \emph {et~al.}(2015)\citenamefont {Embon},
  \citenamefont {Anahory}, \citenamefont {Suhov}, \citenamefont {Halbertal},
  \citenamefont {Cuppens}, \citenamefont {Yakovenko}, \citenamefont {Uri},
  \citenamefont {Myasoedov}, \citenamefont {Rappaport}, \citenamefont {Huber},
  \citenamefont {Gurevich},\ and\ \citenamefont {Zeldov}}]{Emb15nsr}%
  \BibitemOpen
  \bibfield  {author} {\bibinfo {author} {\bibfnamefont {L.}~\bibnamefont
  {Embon}}, \bibinfo {author} {\bibfnamefont {Y.}~\bibnamefont {Anahory}},
  \bibinfo {author} {\bibfnamefont {A.}~\bibnamefont {Suhov}}, \bibinfo
  {author} {\bibfnamefont {D.}~\bibnamefont {Halbertal}}, \bibinfo {author}
  {\bibfnamefont {J.}~\bibnamefont {Cuppens}}, \bibinfo {author} {\bibfnamefont
  {A.}~\bibnamefont {Yakovenko}}, \bibinfo {author} {\bibfnamefont
  {A.}~\bibnamefont {Uri}}, \bibinfo {author} {\bibfnamefont {Y.}~\bibnamefont
  {Myasoedov}}, \bibinfo {author} {\bibfnamefont {M.~L.}\ \bibnamefont
  {Rappaport}}, \bibinfo {author} {\bibfnamefont {M.~E.}\ \bibnamefont
  {Huber}}, \bibinfo {author} {\bibfnamefont {A.}~\bibnamefont {Gurevich}}, \
  and\ \bibinfo {author} {\bibfnamefont {E.}~\bibnamefont {Zeldov}},\ }\href
  {\doibase 10.1038/srep07598} {\bibfield  {journal} {\bibinfo  {journal} {Sci
  Rep.}\ }\textbf {\bibinfo {volume} {5}},\ \bibinfo {pages} {7598} (\bibinfo
  {year} {2015})}\BibitemShut {NoStop}%
\bibitem [{\citenamefont {Kremen}\ \emph {et~al.}(2016)\citenamefont {Kremen},
  \citenamefont {Wissberg}, \citenamefont {Haham}, \citenamefont {Persky},
  \citenamefont {Frenkel},\ and\ \citenamefont {Kalisky}}]{Kre16nal}%
  \BibitemOpen
  \bibfield  {author} {\bibinfo {author} {\bibfnamefont {A.}~\bibnamefont
  {Kremen}}, \bibinfo {author} {\bibfnamefont {S.}~\bibnamefont {Wissberg}},
  \bibinfo {author} {\bibfnamefont {N.}~\bibnamefont {Haham}}, \bibinfo
  {author} {\bibfnamefont {E.}~\bibnamefont {Persky}}, \bibinfo {author}
  {\bibfnamefont {Y.}~\bibnamefont {Frenkel}}, \ and\ \bibinfo {author}
  {\bibfnamefont {B.}~\bibnamefont {Kalisky}},\ }\href {\doibase
  10.1021/acs.nanolett.5b04444} {\bibfield  {journal} {\bibinfo  {journal}
  {Nano Letters}\ }\textbf {\bibinfo {volume} {0}},\ \bibinfo {pages} {null}
  (\bibinfo {year} {2016})},\ \bibinfo {note} {pMID: 26836018}\BibitemShut
  {NoStop}%
\bibitem [{\citenamefont {Lu}\ \emph {et~al.}(2007)\citenamefont {Lu},
  \citenamefont {Reichhardt},\ and\ \citenamefont {Reichhardt}}]{Luq07prb}%
  \BibitemOpen
  \bibfield  {author} {\bibinfo {author} {\bibfnamefont {Q.}~\bibnamefont
  {Lu}}, \bibinfo {author} {\bibfnamefont {C.~J.~O.}\ \bibnamefont
  {Reichhardt}}, \ and\ \bibinfo {author} {\bibfnamefont {C.}~\bibnamefont
  {Reichhardt}},\ }\href {\doibase 10.1103/PhysRevB.75.054502} {\bibfield
  {journal} {\bibinfo  {journal} {Phys. Rev. B}\ }\textbf {\bibinfo {volume}
  {75}},\ \bibinfo {pages} {054502} (\bibinfo {year} {2007})}\BibitemShut
  {NoStop}%
\bibitem [{\citenamefont {Gittleman}\ and\ \citenamefont
  {Rosenblum}(1966)}]{Git66prl}%
  \BibitemOpen
  \bibfield  {author} {\bibinfo {author} {\bibfnamefont {J.~I.}\ \bibnamefont
  {Gittleman}}\ and\ \bibinfo {author} {\bibfnamefont {B.}~\bibnamefont
  {Rosenblum}},\ }\href {\doibase 10.1103/PhysRevLett.16.734} {\bibfield
  {journal} {\bibinfo  {journal} {Phys. Rev. Lett.}\ }\textbf {\bibinfo
  {volume} {16}},\ \bibinfo {pages} {734} (\bibinfo {year} {1966})}\BibitemShut
  {NoStop}%
\bibitem [{\citenamefont {Coffey}\ and\ \citenamefont {Clem}(1991)}]{Cof91prl}%
  \BibitemOpen
  \bibfield  {author} {\bibinfo {author} {\bibfnamefont {M.~W.}\ \bibnamefont
  {Coffey}}\ and\ \bibinfo {author} {\bibfnamefont {J.~R.}\ \bibnamefont
  {Clem}},\ }\href {\doibase 10.1103/PhysRevLett.67.386} {\bibfield  {journal}
  {\bibinfo  {journal} {Phys. Rev. Lett.}\ }\textbf {\bibinfo {volume} {67}},\
  \bibinfo {pages} {386} (\bibinfo {year} {1991})}\BibitemShut {NoStop}%
\bibitem [{\citenamefont {Pompeo}\ and\ \citenamefont
  {Silva}(2008)}]{Pom08prb}%
  \BibitemOpen
  \bibfield  {author} {\bibinfo {author} {\bibfnamefont {N.}~\bibnamefont
  {Pompeo}}\ and\ \bibinfo {author} {\bibfnamefont {E.}~\bibnamefont {Silva}},\
  }\href {\doibase 10.1103/PhysRevB.78.094503} {\bibfield  {journal} {\bibinfo
  {journal} {Phys. Rev. B}\ }\textbf {\bibinfo {volume} {78}},\ \bibinfo
  {pages} {094503} (\bibinfo {year} {2008})}\BibitemShut {NoStop}%
\bibitem [{\citenamefont {Shklovskij}\ and\ \citenamefont
  {Dobrovolskiy}(2008)}]{Shk08prb}%
  \BibitemOpen
  \bibfield  {author} {\bibinfo {author} {\bibfnamefont {V.~A.}\ \bibnamefont
  {Shklovskij}}\ and\ \bibinfo {author} {\bibfnamefont {O.~V.}\ \bibnamefont
  {Dobrovolskiy}},\ }\href {\doibase 10.1103/PhysRevB.78.104526} {\bibfield
  {journal} {\bibinfo  {journal} {Phys. Rev. B}\ }\textbf {\bibinfo {volume}
  {78}},\ \bibinfo {pages} {104526} (\bibinfo {year} {2008})}\BibitemShut
  {NoStop}%
\bibitem [{\citenamefont {Shklovskij}\ and\ \citenamefont
  {Dobrovolskiy}(2011)}]{Shk11prb}%
  \BibitemOpen
  \bibfield  {author} {\bibinfo {author} {\bibfnamefont {V.~A.}\ \bibnamefont
  {Shklovskij}}\ and\ \bibinfo {author} {\bibfnamefont {O.~V.}\ \bibnamefont
  {Dobrovolskiy}},\ }\href {\doibase 10.1103/PhysRevB.84.054515} {\bibfield
  {journal} {\bibinfo  {journal} {Phys. Rev. B}\ }\textbf {\bibinfo {volume}
  {84}},\ \bibinfo {pages} {054515} (\bibinfo {year} {2011})}\BibitemShut
  {NoStop}%
\bibitem [{\citenamefont {Shklovskij}\ and\ \citenamefont
  {Dobrovolskiy}(2012)}]{Shk12inb}%
  \BibitemOpen
  \bibfield  {author} {\bibinfo {author} {\bibfnamefont {V.~A.}\ \bibnamefont
  {Shklovskij}}\ and\ \bibinfo {author} {\bibfnamefont {O.~V.}\ \bibnamefont
  {Dobrovolskiy}},\ }\enquote {\bibinfo {title} {Microwave absorption by
  vortices in superconductors with a washboard pinning potential},}\ in\
  \href@noop {} {\emph {\bibinfo {booktitle} {Superconductors -- Materials,
  Properties and Applications}}},\ \bibinfo {editor} {edited by\ \bibinfo
  {editor} {\bibfnamefont {A.}~\bibnamefont {Gabovich}}}\ (\bibinfo
  {publisher} {InTech, Rijeka},\ \bibinfo {year} {2012})\ Chap.~\bibinfo
  {chapter} {11}, pp.\ \bibinfo {pages} {263--288}\BibitemShut {NoStop}%
\bibitem [{\citenamefont {Shklovskij}\ \emph {et~al.}(2014)\citenamefont
  {Shklovskij}, \citenamefont {Sosedkin},\ and\ \citenamefont
  {Dobrovolskiy}}]{Shk14pcm}%
  \BibitemOpen
  \bibfield  {author} {\bibinfo {author} {\bibfnamefont {V.~A.}\ \bibnamefont
  {Shklovskij}}, \bibinfo {author} {\bibfnamefont {V.~V.}\ \bibnamefont
  {Sosedkin}}, \ and\ \bibinfo {author} {\bibfnamefont {O.~V.}\ \bibnamefont
  {Dobrovolskiy}},\ }\href {http://stacks.iop.org/0953-8984/26/i=2/a=025703}
  {\bibfield  {journal} {\bibinfo  {journal} {J. Phys.: Cond. Matt.}\ }\textbf
  {\bibinfo {volume} {26}},\ \bibinfo {pages} {025703} (\bibinfo {year}
  {2014})}\BibitemShut {NoStop}%
\bibitem [{\citenamefont {Shklovskij}\ and\ \citenamefont
  {Hop}(2010)}]{Shk10ltp}%
  \BibitemOpen
  \bibfield  {author} {\bibinfo {author} {\bibfnamefont {V.~A.}\ \bibnamefont
  {Shklovskij}}\ and\ \bibinfo {author} {\bibfnamefont {D.~T.~B.}\ \bibnamefont
  {Hop}},\ }\href {\doibase 10.1063/1.329293} {\bibfield  {journal} {\bibinfo
  {journal} {Low Temp. Phys.}\ }\textbf {\bibinfo {volume} {36}},\ \bibinfo
  {pages} {71} (\bibinfo {year} {2010})}\BibitemShut {NoStop}%
\bibitem [{\citenamefont {Fiory}(1971)}]{Fio71prl}%
  \BibitemOpen
  \bibfield  {author} {\bibinfo {author} {\bibfnamefont {A.~T.}\ \bibnamefont
  {Fiory}},\ }\href {\doibase 10.1103/PhysRevLett.27.501} {\bibfield  {journal}
  {\bibinfo  {journal} {Phys. Rev. Lett.}\ }\textbf {\bibinfo {volume} {27}},\
  \bibinfo {pages} {501} (\bibinfo {year} {1971})}\BibitemShut {NoStop}%
\bibitem [{\citenamefont {Larkin}\ and\ \citenamefont
  {Ovchinnikov}(1975)}]{Lar75etp}%
  \BibitemOpen
  \bibfield  {author} {\bibinfo {author} {\bibfnamefont {A.~I.}\ \bibnamefont
  {Larkin}}\ and\ \bibinfo {author} {\bibfnamefont {Y.~N.}\ \bibnamefont
  {Ovchinnikov}},\ }\href
  {http://www.jetp.ac.ru/cgi-bin/index/e/41/5/p960?a=list} {\bibfield
  {journal} {\bibinfo  {journal} {J. Exp. Theor. Phys.}\ }\textbf {\bibinfo
  {volume} {41}},\ \bibinfo {pages} {960} (\bibinfo {year} {1975})}\BibitemShut
  {NoStop}%
\bibitem [{\citenamefont {Musienko}\ \emph {et~al.}(1980)\citenamefont
  {Musienko}, \citenamefont {Dmitrenko},\ and\ \citenamefont
  {Volotskaya}}]{Mus80etp}%
  \BibitemOpen
  \bibfield  {author} {\bibinfo {author} {\bibfnamefont {L.~E.}\ \bibnamefont
  {Musienko}}, \bibinfo {author} {\bibfnamefont {I.~M.}\ \bibnamefont
  {Dmitrenko}}, \ and\ \bibinfo {author} {\bibfnamefont {V.~G.}\ \bibnamefont
  {Volotskaya}},\ }\href@noop {} {\bibfield  {journal} {\bibinfo  {journal}
  {JETP Lett.}\ }\textbf {\bibinfo {volume} {31}},\ \bibinfo {pages} {567}
  (\bibinfo {year} {1980})}\BibitemShut {NoStop}%
\bibitem [{\citenamefont {Klein}\ \emph {et~al.}(1985)\citenamefont {Klein},
  \citenamefont {Huebener}, \citenamefont {Gauss},\ and\ \citenamefont
  {Parisi}}]{Kle85ltp}%
  \BibitemOpen
  \bibfield  {author} {\bibinfo {author} {\bibfnamefont {W.}~\bibnamefont
  {Klein}}, \bibinfo {author} {\bibfnamefont {R.~P.}\ \bibnamefont {Huebener}},
  \bibinfo {author} {\bibfnamefont {S.}~\bibnamefont {Gauss}}, \ and\ \bibinfo
  {author} {\bibfnamefont {J.}~\bibnamefont {Parisi}},\ }\href {\doibase
  10.1007/BF00683694} {\bibfield  {journal} {\bibinfo  {journal} {J. Low Temp.
  Phys.}\ }\textbf {\bibinfo {volume} {61}},\ \bibinfo {pages} {413} (\bibinfo
  {year} {1985})}\BibitemShut {NoStop}%
\bibitem [{\citenamefont {Volotskaya}\ \emph {et~al.}(1992)\citenamefont
  {Volotskaya}, \citenamefont {Dmitrenko}, \citenamefont {Koretskaya},\ and\
  \citenamefont {Musienko}}]{Vol92fnt}%
  \BibitemOpen
  \bibfield  {author} {\bibinfo {author} {\bibfnamefont {V.~G.}\ \bibnamefont
  {Volotskaya}}, \bibinfo {author} {\bibfnamefont {I.~M.}\ \bibnamefont
  {Dmitrenko}}, \bibinfo {author} {\bibfnamefont {O.~A.}\ \bibnamefont
  {Koretskaya}}, \ and\ \bibinfo {author} {\bibfnamefont {L.~E.}\ \bibnamefont
  {Musienko}},\ }\href@noop {} {\bibfield  {journal} {\bibinfo  {journal} {Fiz.
  Nizk. Temp.}\ }\textbf {\bibinfo {volume} {18}},\ \bibinfo {pages} {973}
  (\bibinfo {year} {1992})}\BibitemShut {NoStop}%
\bibitem [{\citenamefont {Peroz}\ and\ \citenamefont
  {Villard}(2005)}]{Per05prb}%
  \BibitemOpen
  \bibfield  {author} {\bibinfo {author} {\bibfnamefont {C.}~\bibnamefont
  {Peroz}}\ and\ \bibinfo {author} {\bibfnamefont {C.}~\bibnamefont
  {Villard}},\ }\href {\doibase 10.1103/PhysRevB.72.014515} {\bibfield
  {journal} {\bibinfo  {journal} {Phys. Rev. B}\ }\textbf {\bibinfo {volume}
  {72}},\ \bibinfo {pages} {014515} (\bibinfo {year} {2005})}\BibitemShut
  {NoStop}%
\bibitem [{\citenamefont {Leo}\ \emph {et~al.}(2016)\citenamefont {Leo},
  \citenamefont {Marra}, \citenamefont {Grimaldi}, \citenamefont {Citro},
  \citenamefont {Kawale}, \citenamefont {Bellingeri}, \citenamefont
  {Ferdeghini}, \citenamefont {Pace},\ and\ \citenamefont {Nigro}}]{Leo16prb}%
  \BibitemOpen
  \bibfield  {author} {\bibinfo {author} {\bibfnamefont {A.}~\bibnamefont
  {Leo}}, \bibinfo {author} {\bibfnamefont {P.}~\bibnamefont {Marra}}, \bibinfo
  {author} {\bibfnamefont {G.}~\bibnamefont {Grimaldi}}, \bibinfo {author}
  {\bibfnamefont {R.}~\bibnamefont {Citro}}, \bibinfo {author} {\bibfnamefont
  {S.}~\bibnamefont {Kawale}}, \bibinfo {author} {\bibfnamefont
  {E.}~\bibnamefont {Bellingeri}}, \bibinfo {author} {\bibfnamefont
  {C.}~\bibnamefont {Ferdeghini}}, \bibinfo {author} {\bibfnamefont
  {S.}~\bibnamefont {Pace}}, \ and\ \bibinfo {author} {\bibfnamefont
  {A.}~\bibnamefont {Nigro}},\ }\href {\doibase 10.1103/PhysRevB.93.054503}
  {\bibfield  {journal} {\bibinfo  {journal} {Phys. Rev. B}\ }\textbf {\bibinfo
  {volume} {93}},\ \bibinfo {pages} {054503} (\bibinfo {year}
  {2016})}\BibitemShut {NoStop}%
\bibitem [{\citenamefont {Knufinke}\ \emph {et~al.}(2012)\citenamefont
  {Knufinke}, \citenamefont {Ilin}, \citenamefont {Siegel}, \citenamefont
  {Koelle}, \citenamefont {Kleiner},\ and\ \citenamefont
  {Goldobin}}]{Knu12pre}%
  \BibitemOpen
  \bibfield  {author} {\bibinfo {author} {\bibfnamefont {M.}~\bibnamefont
  {Knufinke}}, \bibinfo {author} {\bibfnamefont {K.}~\bibnamefont {Ilin}},
  \bibinfo {author} {\bibfnamefont {M.}~\bibnamefont {Siegel}}, \bibinfo
  {author} {\bibfnamefont {D.}~\bibnamefont {Koelle}}, \bibinfo {author}
  {\bibfnamefont {R.}~\bibnamefont {Kleiner}}, \ and\ \bibinfo {author}
  {\bibfnamefont {E.}~\bibnamefont {Goldobin}},\ }\href {\doibase
  10.1103/PhysRevE.85.011122} {\bibfield  {journal} {\bibinfo  {journal} {Phys.
  Rev. E}\ }\textbf {\bibinfo {volume} {85}},\ \bibinfo {pages} {011122}
  (\bibinfo {year} {2012})}\BibitemShut {NoStop}%
\bibitem [{\citenamefont {Shklovskij}(2008)}]{Shk08mmt}%
  \BibitemOpen
  \bibfield  {author} {\bibinfo {author} {\bibfnamefont {V.~A.}\ \bibnamefont
  {Shklovskij}},\ }in\ \href@noop {} {\emph {\bibinfo {booktitle} {Procedings
  of the Fifth International Conference on Mathematical Modeling and Computer
  Simulation of Materials Technologies MMT-2008, Ariel, Israel}}}\ (\bibinfo
  {year} {2008})\BibitemShut {NoStop}%
\bibitem [{\citenamefont {Shklovskij}\ and\ \citenamefont
  {Dobrovolskiy}(2013)}]{Shk13ltp}%
  \BibitemOpen
  \bibfield  {author} {\bibinfo {author} {\bibfnamefont {V.~A.}\ \bibnamefont
  {Shklovskij}}\ and\ \bibinfo {author} {\bibfnamefont {O.~V.}\ \bibnamefont
  {Dobrovolskiy}},\ }\href {\doibase 10.1063/1.4791773} {\bibfield  {journal}
  {\bibinfo  {journal} {Low Temp. Phys.}\ }\textbf {\bibinfo {volume} {39}},\
  \bibinfo {pages} {120} (\bibinfo {year} {2013})}\BibitemShut {NoStop}%
\bibitem [{\citenamefont {Dobrovolskiy}\ \emph {et~al.}(2012)\citenamefont
  {Dobrovolskiy}, \citenamefont {Begun}, \citenamefont {Huth},\ and\
  \citenamefont {Shklovskij}}]{Dob12njp}%
  \BibitemOpen
  \bibfield  {author} {\bibinfo {author} {\bibfnamefont {O.~V.}\ \bibnamefont
  {Dobrovolskiy}}, \bibinfo {author} {\bibfnamefont {E.}~\bibnamefont {Begun}},
  \bibinfo {author} {\bibfnamefont {M.}~\bibnamefont {Huth}}, \ and\ \bibinfo
  {author} {\bibfnamefont {V.~A.}\ \bibnamefont {Shklovskij}},\ }\href
  {http://stacks.iop.org/1367-2630/14/i=11/a=113027} {\bibfield  {journal}
  {\bibinfo  {journal} {New J. Phys.}\ }\textbf {\bibinfo {volume} {14}},\
  \bibinfo {pages} {113027} (\bibinfo {year} {2012})}\BibitemShut {NoStop}%
\bibitem [{\citenamefont {Dobrovolskiy}\ and\ \citenamefont
  {Huth}(2012)}]{Dob12tsf}%
  \BibitemOpen
  \bibfield  {author} {\bibinfo {author} {\bibfnamefont {O.~V.}\ \bibnamefont
  {Dobrovolskiy}}\ and\ \bibinfo {author} {\bibfnamefont {M.}~\bibnamefont
  {Huth}},\ }\href {\doibase 10.1016/j.tsf.2012.04.083} {\bibfield  {journal}
  {\bibinfo  {journal} {Thin Solid Films}\ }\textbf {\bibinfo {volume} {520}},\
  \bibinfo {pages} {5985} (\bibinfo {year} {2012})}\BibitemShut {NoStop}%
\bibitem [{\citenamefont {Dobrovolskiy}\ \emph
  {et~al.}(2015{\natexlab{b}})\citenamefont {Dobrovolskiy}, \citenamefont
  {Franke},\ and\ \citenamefont {Huth}}]{Dob15mst}%
  \BibitemOpen
  \bibfield  {author} {\bibinfo {author} {\bibfnamefont {O.~V.}\ \bibnamefont
  {Dobrovolskiy}}, \bibinfo {author} {\bibfnamefont {J.}~\bibnamefont
  {Franke}}, \ and\ \bibinfo {author} {\bibfnamefont {M.}~\bibnamefont
  {Huth}},\ }\href {http://stacks.iop.org/0957-0233/26/i=3/a=035502} {\bibfield
   {journal} {\bibinfo  {journal} {Meas. Sci. Technol.}\ }\textbf {\bibinfo
  {volume} {26}},\ \bibinfo {pages} {035502} (\bibinfo {year}
  {2015}{\natexlab{b}})}\BibitemShut {NoStop}%
\end{thebibliography}

%

\end{document}